\documentclass[12pt,preprint]{aastex}
\usepackage{epstopdf}
\usepackage{txfonts}
\usepackage{natbib}
\shorttitle{Structure of the Small Magellanic Cloud}
\shortauthors{S. Subramanian \& A. Subramaniam}
\begin{document}
\title{The 3D structure of the Small Magellanic Cloud}
\author{Smitha Subramanian\altaffilmark{1,2},
Annapurni Subramaniam\altaffilmark{1}}
\affil{Indian Institute of Astrophysics, Koramangala II Block, Bangalore - 34, India\\
Department of Physics, Calicut University, Calicut, Kerala, India\\}
\begin{abstract}
The 3D structure of the inner Small Magellanic Cloud (SMC) is investigated using the red clump (RC) 
stars and the RR Lyrae stars (RRLS), which represent the intermediate-age and the old stellar 
populations of a galaxy. The V and I bands photometric data from the Optical Gravitational Lensing 
Experiment (OGLE III) catalog are used for our study. The mean dereddened I${_0}$ magnitude of the RC stars
and the RRLS are used to study the relative positions of the different regions in the SMC with respect 
to the mean SMC distance. This shows that the northeastern part of the SMC is closer to us. The line of 
sight depth (front to back distance) across the SMC is estimated using the dispersion in the I${_0}$ magnitudes 
of both the RC stars and the RRLS and found to be large ($\sim$ 14 kpc) for both the populations. 
The similarity in their depth distribution suggest that both these 
populations occupy a similar volume of the SMC. 
The surface density distribution and the radial density profile of the RC stars 
suggest that they are more likely to be distributed in a nearly spheroidal 
system. The tidal radius estimated for the SMC system is $\sim$ 7-12 kpc.
An elongation along the NE$-$SW direction is seen in the surface density map of the RC stars. 
The surface density distribution of the RRLS in the SMC is nearly circular. Based on all the 
above results the observed structure of the SMC, in which both the RC stars and RRLS are distributed, 
is approximated as a triaxial ellipsoid. The parameters of the ellipsoid are obtained 
using the inertia tensor analysis. 
We estimated the axes ratio, inclination of the longest axis with 
the line of sight ($\it{i}$) and the position angle ($\phi$) of the longest axis of the elliposid on 
the sky from the analysis of the RRLS. The analysis of the red clump stars with the assumption that 
they are extended up to a depth of 3.5 times the sigma (width of dereddened I$_0$ magnitude distribution, 
corrected for intrinsic spread and observational errors), was also found to give similar axes ratio 
and orientation angles. The above estimated parameters depend on the data coverage of 
the SMC. Using the RRLS with equal coverage in all the three axes (data within 3 
degrees in X, Y and Z axes), we estimated an axes ratio of 1:1.33:1.61 with  
$\it{i}$ = 2$^\circ$.6  and $\phi$ = 70$^\circ$.2. 
Our tidal radius estimates and the recent observational studies suggest 
that the full extent of the SMC in the XY plane is of the order of the 
front to back distance estimated along the line of sight. These results suggest that 
the structure of the SMC  is spheroidal or slightly ellipsoidal.
We propose that the SMC experienced a merger with 
another dwarf galaxy at $\sim$ 4$-$5 Gyr ago , and the merger process was completed in 
another 2$-$3 Gyr. This resulted in a spheroidal distribution comprising of stars 
older than 2 Gyr. 
\end{abstract}

\keywords{Magellanic Clouds, galaxies: structure, Stars: variable stars, horizontal branch stars}

\section{Introduction}
The Small Magellanic Cloud (SMC) located at a distance of around 60 kpc is one 
of the nearest galaxies. The SMC is characterized by a less pronounced bar,  
than that seen in the Large Magellanic Cloud (LMC). It also has an eastern extension called the Wing. 
The Wing and the northeastern 
part of the bar are closer than the southern parts \citep{h93}.
A large line of sight depth was found in the outer and inner regions of the 
SMC by \cite{gh91} and  \cite{ss09} respectively. 

In an extensive survey of variable stars in the SMC, \cite{g75} discovered 76 RR Lyrae stars (RRLS) in the 
field centered on the globular cluster, NGC 121. He found that the period distribution of the RRLS 
in the SMC is unlike that found anywhere in the Galaxy and closely resembles the 
distribution of variable stars in the Leo II dwarf galaxy. 
His studies also showed that these stars are  
distributed rather evenly and concentrate neither towards the bar nor towards the 
center of the SMC.
He mentioned that the lack of strong stellar 
concentration is another property in common with the stellar populations in the 
dwarf spheroidal galaxies. \cite{s92} and 
\cite{s02} also 
found an even and smooth distribution of RRLS in the northeastern regions and the central 
2.4 square degree regions of the SMC respectively.  \cite{ss09} used the 
RRLS data published by 
\cite{s02} to estimate the depth in the central regions of 
the SMC. They found that the depth estimated using the 
red clump (RC) stars and the RRLS are similar in the central regions of the SMC. 
They suggested that the RRLS and the RC stars 
in the central bar region of the SMC are born in the same location and occupy a
similar volume in the galaxy. 
\cite{s10} presented a catalog of the 
RRLS in the SMC from Optical Gravitational Lensing Experiment (OGLE) III survey. From the 
spatial distribution of the RRLS they suggested that the halo of the SMC is roughly 
circular in the sky. However their density map of the RRLS revealed two maxima near the center of the 
SMC. 

\cite{z00} showed that the older stellar populations (age $>$ 1 Gyr) 
in the SMC are distributed in a regular, smooth ellipsoid. Similar conclusions 
were drawn by \cite{c00} from the DENIS near infrared survey. 
\cite{m01} further investigated the dynamical origin of the 
bar using isodensity contour maps of stars with different ages. 
They found similar results for old stellar populations. 
A sample of 12 populous SMC clusters which possess RC stars are studied by 
\cite{c01} to determine the distances to these clusters. The line of sight depth 
of the SMC is estimated as the standard deviation (sigma) of these distances.   
They found a 1-sigma depth of $\approx$ 6-12 kpc for the SMC. Viewing the 
SMC as a triaxial ellipsoid with RA, DEC and the line of sight depth as the 
three axes, they found an axes ratio of 1:2:4. From a spectroscopic study of 2046 
red giant stars \cite{hz06} found that the older stellar 
components of the SMC have a velocity dispersion of 27.5 kms$^{-1}$ and a 
maximum possible rotation of 17 kms$^{-1}$. Their result is consistent with other kinematical 
studies based on the radial velocities of the PNe and carbon stars, which 
represent the old and intermediate-age stellar populations (\citealt{d85}, 
\citealt{s86} \& \citealt{h97}). This implies that 
the structure of the older stellar component of the SMC is a spheroidal/ellipsoidal, 
that is supported by its velocity dispersion.

From the analysis of young stars (age $<$ 200 Myr) \cite{z00} suggested 
that the irregular appearance of the SMC is due to the recent star formation. 
As in the case of young stars, the large scale HI morphology of the SMC 
obtained from the high resolution HI observations \citep{s04} is also 
quite irregular and does not show symmetry.
The most prominent features are the elongation from the northeast 
to the southwest and the V-shaped concentration in the east. 
The HI observations also show that the SMC has a significant amount of rotation 
with a circular velocity of approximately 60 kms$^{-1}$ \citep{s04} and a 
large velocity gradient of 91 kms$^{-1}$ in the southwest to 200 kms$^{-1}$ in the north 
east. \citep{eh08} obtained velocities for 2045 young (O, B, A) stars
in the SMC, and found a velocity gradient of similar slope as seen in the HI gas. 
Surprisingly though, they found a position angle  ($\sim$ 126$^{\circ}$)  for the line of 
maximum velocity gradient that is quite different, and almost orthogonal to that seen 
in the HI. \cite{v09} suggested that this difference in the position 
angles may be an artifact of the different spatial coverage of the two studies 
(\cite{eh08} did not observe in the North-East region where the HI velocities 
are the largest), since it would be difficult to find a physical explanation for a 
significant difference in kinematics between HI gas and young stars. The inclination 
of the SMC disk in which the young stars are believed to be distributed 
is estimated as 70$^{\circ}$ $\pm$ 3$^{\circ}$ and 68$^{\circ}$ $\pm$ 2$^{\circ}$ from the photometric   
studies of Cepheids by \cite{cc86} and \cite{g00} respectively. The position 
angle of the line of nodes is estimated to be $\sim$ 148$^{\circ}$.0. 

\cite{gn96} modelled the SMC as a two component system consisting of 
a nearly spherical halo and a rotationally supported disc. The tidal radius of the 
SMC is estimated as 5 kpc according to their model. Their best fit to the distribution 
halo particles within 5 kpc radius was in good agreement with the observed 
distributions of the old ($>$ 9 Gyr) and intermediate age (2$-$9 Gyr)
stellar populations in the SMC. The distribution of disc particles could reproduce 
the observed irregular distribution of young stars in the SMC. 
\cite{bk08b} suggested that a major merger event in the early 
stage of the SMC formation caused the coexistence of a spheroidal stellar component 
and an extended rotating HI disk.

Both the observational and theoretical studies suggest that the old and the intermediate-age 
stellar populations in the SMC are distributed in a spheroidal/ellipsoidal component. 
Motivated with this result, here we study the RC stars and the 
RRLS in the inner SMC, which represent the intermediate-age and the old 
stellar populations respectively. This study aims to understand the structure of the 
inner SMC and hence to quantitatively estimate the structural parameters.
 
The RC stars are core helium burning 
stars, which are metal rich and slightly more massive counter parts of the horizontal branch 
stars. They have tightly defined colour and magnitude, and appear as an easily identifiable 
component in the colour magnitude diagrams (CMDs). These properties of the RC stars make them 
a good proxy to determine the distance to a galaxy and also to understand the structure 
of a galaxy. Previously many people have used the RC stars as a proxy for distance estimation  
\citep{s98}, to identify the structures in the LMC \citep{s03} and to estimate the orientation measurements of the LMC (\citealt{os02} 
and \citealt{ss10}). \cite{ss09} estimated the depth 
of both the LMC \& the SMC using the dispersions in magnitude and colour distributions of the RC 
stars.

The RRLS are metal poor, low mass, core helium burning stars which undergo radial 
pulsations. Their period of pulsation have a range of 0.2-1 day. The RRLS are excellent 
tracers of the oldest observable population of stars in a galaxy. 
The RRab type stars have constant mean magnitude and they are 
used for distance estimation 
\citep{b09}. The dispersion about the mean magnitude is 
a measure of the depth of the host galaxy. \cite{s06} and 
\cite{ss09}  used the dispersion of the RRab  
stars to estimate the line of sight depth of the LMC and SMC respectively. 
\cite{ps09} studied the structure of the 
LMC stellar halo using the RRLS in the LMC.  

In this paper we estimate the structural parameters of the older component of the 
SMC from the analysis of the RC stars and the RRLS in the inner SMC.  
\cite{u08} presented the V and I bands photometric data of the 16 sq.degrees 
of the SMC from the OGLE III survey. The catalog of the SMC RRLS from the 
OGLE III survey is presented by \cite{s10}. Both these data sets are used in this 
study. The mean dereddened I$_0$ magnitude of the RC stars and the RRLS are used 
to estimate the relative locations of different regions in the SMC with respect 
to the mean SMC distance. The dispersion in the mean magnitude is used to 
obtain the line of sight depth across the SMC.

The structure of the paper is as follows. 
The next section deals with the contribution of heterogeneous population on 
the dereddened I$_0$ magnitude of the RC stars and RRLS and also to
the dispersion in their mean magnitude.  
In section 3 we explain 
the data and analysis of both the RC stars and the RRLS. The reddening map of the SMC is 
presented in section 4. The results of our analysis are 
given in section 5. The estimation of the structural parameters of the SMC are explained in 
detail in section 6. The discussion and the conclusions are presented in section 7 \& 8 
respectively.

\section{Contribution of heterogeneous population}

\cite{s98} determined 
the distance to the Magellanic Clouds (MCs) as well as to the bulge of our Galaxy 
from the dereddened apparent magnitude of the RC stars assuming a constant 
absolute I band magnitude for them. Later 
\cite{Sar99} and \cite{C98} claimed that the luminosity 
of the RC stars is highly influenced by the age and metallicity 
and must be accounted for in the determination of the distance. When the 
controversy regarding the use of RC stars as the absolute 
indicator continued, they were used  to estimate relative 
distances between regions within MCs. \cite{os02}, \cite{s03} and \cite{ss10} 
used them as relative distance indicators in their studies. 

The RC stars in the SMC are a heterogeneous population and hence, 
they would have a range in mass, age and metallicity. The density of stars 
in various location will also vary with the local star formation rate as a 
function of time. These factors would contribute to the range of magnitude
and colour of the net population of the RC stars in any given location 
of the SMC and hence to the observed peak magnitude and dispersion. 
\cite{GS01} simulated the RC stars in SMC
using star formation history and age metallicity relation from \cite{PT98}. They 
created the synthetic CMDs of the SMC  and 
the distributions of the RC stars were fitted using numerical analysis to obtain 
the mean and dispersion of the magnitude and colour distributions. 
The model predicted values are used in our study to account for the population effects 
of the RC stars. The estimated intrinsic widths of the colour and magnitude 
distributions of the RC stars in the SMC are 0.03 mag and 0.076 mag respectively. 

The RRLS in the SMC are also a heterogeneous population with a range 
in age, metallicity and mass. There are small variations in their 
luminosities due to evolutionary effects. These factors could contribute 
to the observed dereddened magnitudes of these stars and also to the 
dispersion of magnitude within a sample. This is discussed in section 
3.2.

Apart from multiple population in a given region, there could be 
variation in population and metallicity across the SMC. 
\cite{C06SMC} did not find any different population or 
metallcity gradient near the central regions of the SMC. \cite{T08} 
obtained deep CMDs of 6 SMC inner regions to study the star formation history. 
The interesting fact they found about the six CMDs is the apparent homogeneity 
of old stellar population, populating the subgiant branch and the clump. 
This suggested that there is no large differences in age and metallicity 
exist among old stars in the inner SMC.

\section{Data and Analysis}

\subsection{Red Clump Stars}

The OGLE III survey \citep{u08} presented the V and I bands
photometric data of 16 deg$^2$ of the SMC consisting of about 35 million
stars. We divided the observed region into 1280 regions with a bin size of
8.88 x 4.44 square arcmin. 
The average photometric error of red clump stars
in the I and V bands are around 0.05 mag. 
Photometric data with errors
less than 0.15 mag are considered for the analysis.
For each sub-region the (V$-$I) vs I colour magnitude diagram (CMD) is
plotted and the RC stars are identified. A sample CMD is shown is figure
1. For all the regions, red clump stars are found to be located well within the box shown in the CMD, with
boundaries 0.65 - 1.35 mag in (V$-$I) colour and 17.5 - 19.5 mag in I
magnitude. The number of the RC stars identified in each sub-region ranges from 100$-$3000. 

\begin{figure}{}
\plotone{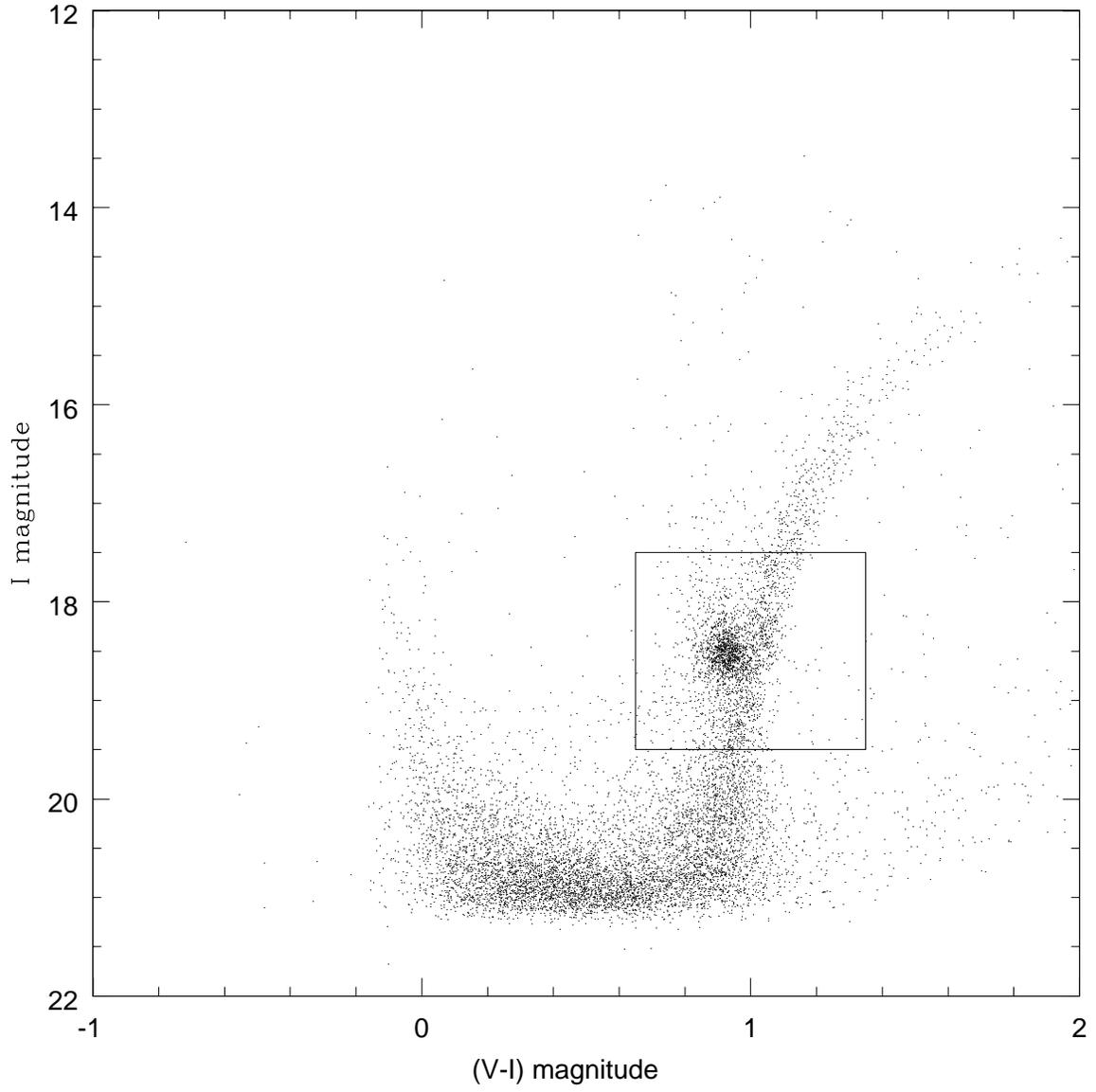}
\caption{The colour magnitude diagram of a sub-region in the SMC. The box used to identify the 
red clump stars is also shown.}
\label{fig1} 
\end{figure}

The RC stars occupy a compact region in the CMD.
Their number distribution profiles resemble a Gaussian. 
The peak values of their colour and magnitude distributions are used to 
obtain the mean dereddened RC magnitude and hence the mean line of 
sight distance to each sub-region in the SMC. The dispersions in the 
magnitude and colour distribution of the RC stars are used to obtain the 
line of sight depth of each sub-region in the SMC.
 
To obtain the number distribution of the RC stars, they are binned in colour 
and magnitude with a bin size of 0.01 and 0.025 mag respectively. These 
distributions are fitted with a Gaussian + Quadratic polynomial. The Gaussian 
represents the RC stars and the other terms represent the red giants in the 
region. A nonlinear least-square method is used to fit the profile and to 
obtain the parameters. The colour and magnitude distributions along with the 
fitted profiles of a sub-region in the SMC are shown in the lower left panels 
of figure 2 and figure 3 respectively. The parameters obtained are the coefficients of each 
term in the function used to fit the profile, error in the estimate of each 
parameter, and reduced $\chi$$^2$ value. For all the sub-regions 
we estimated the peaks and the widths in the I mag and 
(V$-$I) mag of the distributions, associated errors with the parameters, 
and reduced $\chi$$^2$ values. The errors associated with the parameters 
are obtained using the covariance matrix, where the square root of the diagonal 
elements of the matrix gives the error values.

The number of the RC stars identified within the box in the CMD are less (100-400) for 
602 sub-regions. These sub-regions are located towards the edge of the survey and in the 
three isolated fields in the north western region of the SMC. 
For these 602 sub-regions we carefully analyzed 
the magnitude and colour distributions. Because of the less number of stars the peaks 
of the distributions were not very clear and unique. It was difficult to fit those 
distributions with the profiles. As our methodology depends very much 
on the statistical analysis of the sample we omitted these 602 sub-regions, with RC stars 
less than 400, for the remaining analysis. Thus the number of regions used for the study  
reduced to 678. The magnitude and colour distributions of most of these 678 regions 
were checked manually and found that the fits are satisfactory.

\begin{figure}
\plotone{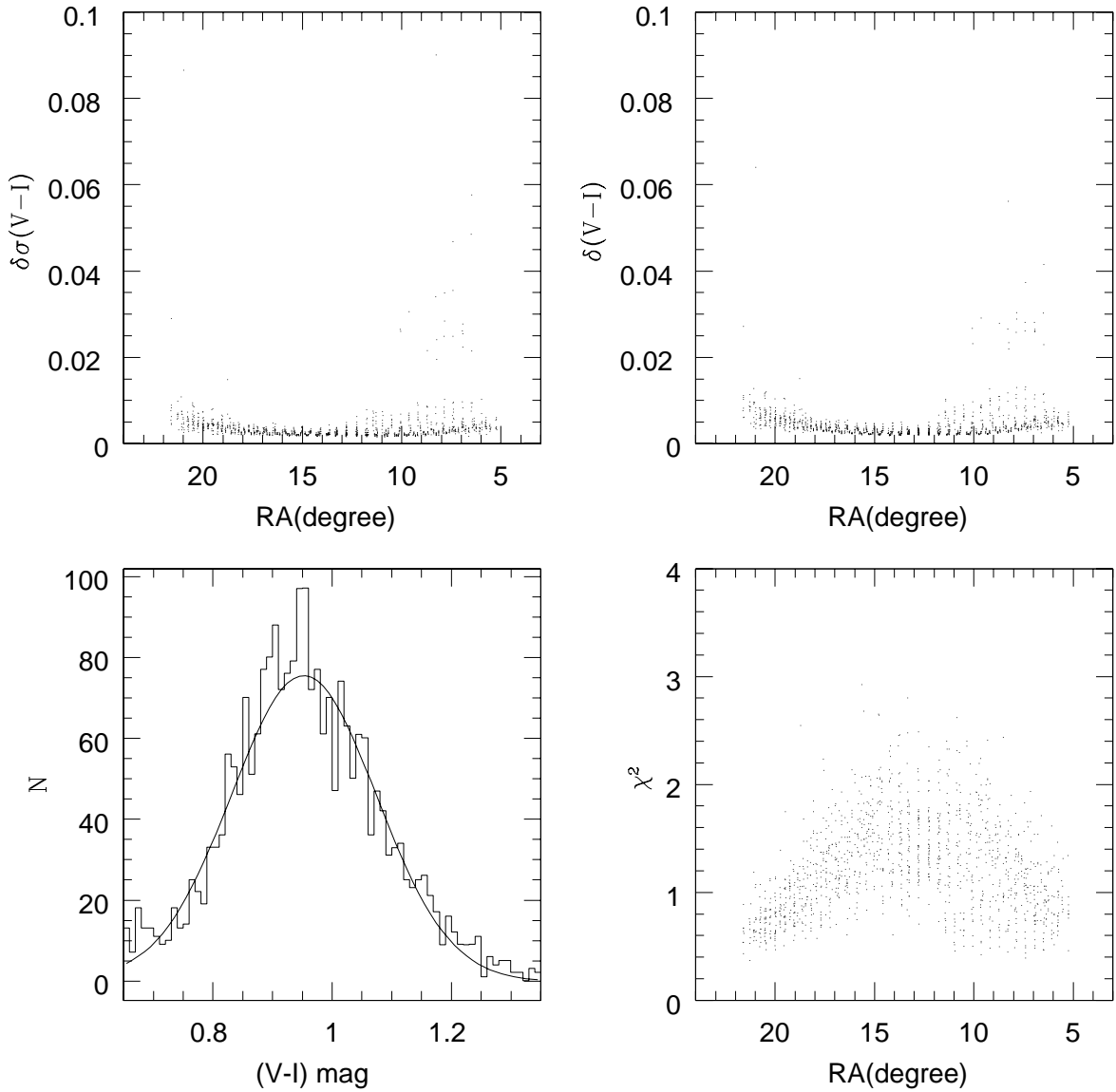}
\caption{A typical colour distribution of a sub-region is shown in the lower left panel.
The best fit to the distribution is also shown. The reduced $\chi$$^2$ values   
and the fit errors in the peak and width of the colour distribution 
of all sub-regions are plotted 
against RA in the lower right, upper right and upper left panels respectively.} 
\label{fig2} 

\end{figure}

\begin{figure}
\plotone{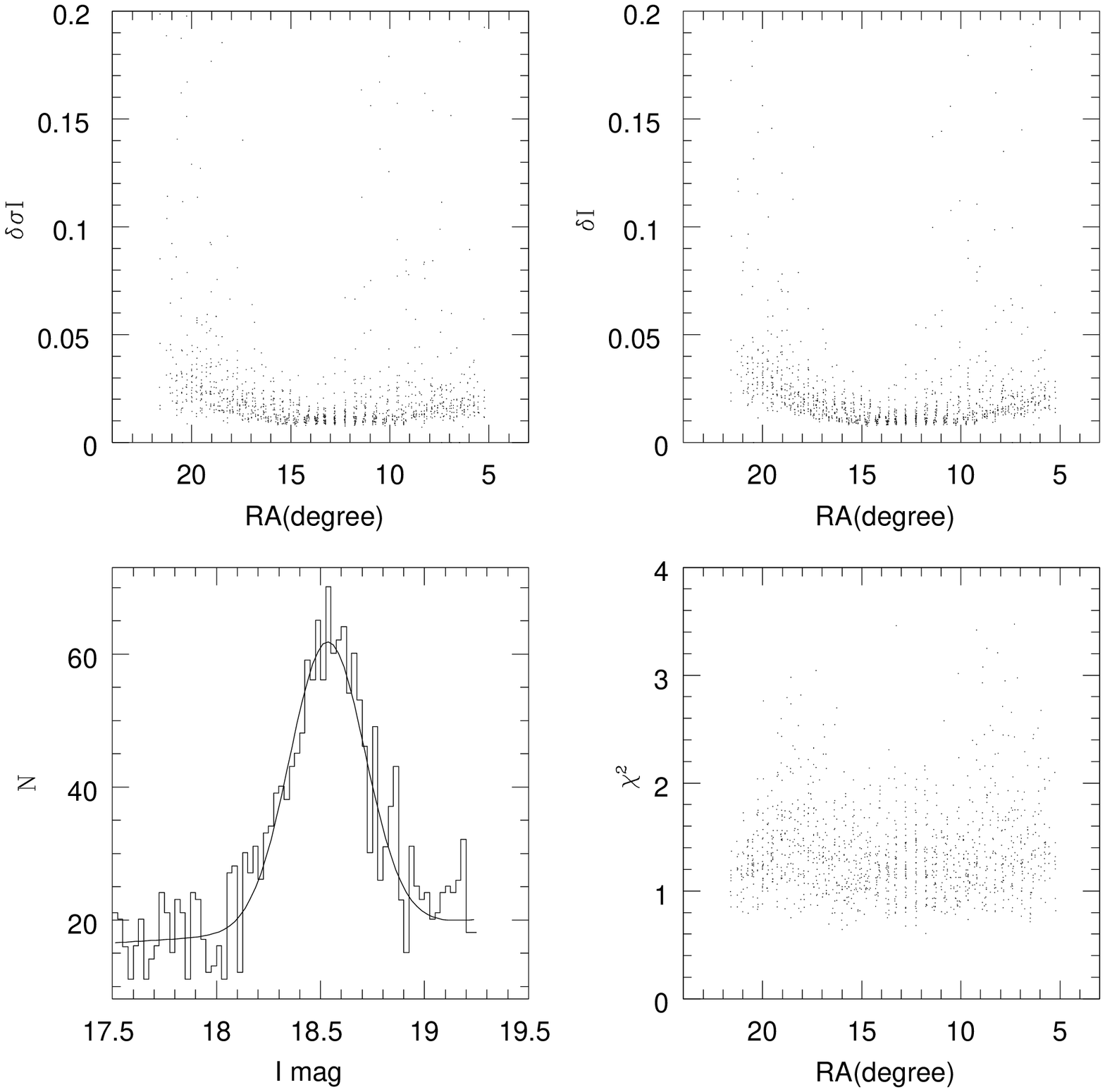}
\caption{A typical magnitude distribution of a sub-region is shown in the lower left panel.
The best fit to the distribution is also shown. The reduced $\chi$$^2$ values   
and the fit errors in the peak and width of the magnitude distribution 
of all the sub-regions are plotted 
against RA in the lower right, upper right and upper left panels respectively.}
\label{fig3} 

\end{figure}

The reduced $\chi$$^2$ values and the fit errors of the peak and width of 
colour and magnitude distributions are plotted against RA in the 
lower right, upper right and upper left panels of figure 2 and figure 3 respectively.  
The sub-regions with reduced $\chi$$^2$ values greater 
than 2.0 for both magnitude and colour distributions are omitted from the 
analysis. The regions with peak and width error greater than 0.05 mag in 
the magnitude distributions are also removed from the analysis. In the 
case of colour distribution, regions with peak and width error greater than 0.02 mag are 
not considered for the analysis. Thus we used only 553 sub-regions out of 
678 sub-regions, for the final analysis. 

The peak value of the colour, (V$-$I) mag at each location is used to 
estimate the reddening. The reddening is calculated using the relation 
E(V$-$I) = (V$-$I)$_{obs}$ - 0.89 mag. The intrinsic colour of the RC stars 
in the SMC is chosen as 0.89 mag to produce a median reddening equal to 
that measured by \cite{sc98} towards the SMC. The interstellar 
extinction is estimated by A${_I}$ = 1.4xE(V$-$I) \citep{s05}. After 
correcting the mean I mag for interstellar extinction, I$_0$ mag for each 
region is estimated. 

The variation in the I${_0}$ mag between the sub-regions is assumed only due to the difference in 
the relative distances. 
The difference in I$_0$ mag is converted into relative distance, $\Delta$D using the distance 
modulus formula,\\
 
(I$_0$ mean $-$ I$_0$ of each region) = 5log$_{10}$(D$_0$/(D$_0$$\pm$$\Delta$D)\\

where D$_0$ is the mean distance to the SMC which is taken as 60kpc.
The average error in I${_0}$ is calculated using the formula, 
$\delta$I${_0}$$^2$ = (avg error in peak I)$^2$ + (1.4 x avg error 
in peak (V$-$I))$^2$, and the error is estimated as 0.013 mag which 
corresponds to $\sim$ 360 pc.

The cartesian coordinates corresponding to each sub-region can be obtained 
using the RA, Dec \& $\Delta$D. The x axis is antiparallel to the RA axis, y axis parallel to 
the declination axis and the z axis towards the observer. The origin of the system is the 
centroid of our sample. The centroid estimated as the mean of RA and Dec of the 
final sample of the RC stars in the 553 sub-regions of the SMC is $\alpha$ =  0$^h$ 52$^m$ 34$^s$.2 $\delta$ = -73$^{\circ}$ 2$^{'}$ 48$^{''}$. 
The x,y, and  z coordinates are obtained using the transformation 
equations given below (\citealt{v01}, see also Appendix of \citealt{wn01}, 
similar transformation for the LMC is given in \citealt{ss10}).\\
\\
x = -Dsin($\alpha$ - $\alpha_0$)cos$\delta$,\\\\
y = Dsin$\delta$cos$\delta_0$ - Dsin$\delta_0$cos($\alpha$ - $\alpha_0$)cos$\delta$,\\\\
z = D$_0$ - Dsin$\delta$sin$\delta_0$ - Dcos$\delta_0$cos($\alpha$ - $\alpha_0$)cos$\delta$,\\

 where D$_0$ is the distance to the SMC and D, the distance to each 
sub-region is given by D = D$_0$ $\pm$ $\Delta$D.
The ($\alpha$, $\delta$) and ($\alpha_0$ , $\delta_0$) represent the RA and Dec of 
each sub-region and the centroid of the sample respectively. 

The estimated dispersions of the colour and magnitude distributions are used 
to obtain the line of sight depth of each region in the SMC.
The total width of the Gaussian in the distribution of colour is due to 
internal reddening, apart from observational error and population effects. 
The width in the distribution of magnitude is due to population effects, 
observational error, internal extinction and depth. By deconvolving the 
effects of observational error, extinction and population effects from the 
distribution of magnitude, an estimate of the depth can be obtained. 
The contribution due to population effects on the observed dispersions 
discussed in section 2 are corrected using the model predicted 
values given by \cite{GS01}. This 
analysis for the estimation of the depth is similar to that done by 
\cite{ss09} for the OGLE II and the Magellanic Cloud 
Photometric Survey (MCPS) regions 
of the SMC.\\ 

$\sigma^2_{col}$ = $\sigma^2_{internal-reddening}$ +
$\sigma$$^2$$_{intrinsic}$ + $\sigma$$^2$$_{error}$\\

$\sigma_{internal-extinction}$ = 1.4 x $\sigma_{internal-reddening}$\\

$\sigma$$^2$$_{mag}$ = $\sigma$$^2$$_{depth}$ +
$\sigma$$^2$$_{internal-extinction}$ +
$\sigma$$^2$$_{intrinsic}$ + $\sigma$$^2$$_{error}$\\

Thus the width corresponding to the depth of the RC distribution in each sub-region 
is estimated. The error in the depth estimate which has contributions from the error in the widths 
of magnitude and colour is also calculated. The average error in the depth estimate is 
$\sim$ 400 pc.\\

\subsection{RR Lyrae stars}
A catalogue of the SMC RRLS from the OGLE III survey is presented 
by 
\cite{s10}. To study the old stellar population in the SMC we 
used 1933 fundamental mode RRLS present in the catalog. We removed 29 
stars out of 1933 which are brighter than the SMC RRLS and are 
possible Galactic objects.

The ab type RRLS could be considered to belong to a similar class and hence 
assumed to have similar properties. The mean magnitude of these stars in the 
I band, after correcting for extinction effects, can be used for 
the estimation of distance and the observed dispersion in their mean 
magnitude is a measure of the depth in their distribution. The reddening 
obtained using the RC stars (described in the previous  
sub-section) is used to estimate the 
extinction to individual RRLS. Stars within each bin of the reddening map are 
assigned a single reddening and it is assumed that reddening does not vary 
much within the bin. Thus the extinction corrected I$_0$ magnitude for all 
the RRLS are estimated. 
As in the case of the RC stars, the difference in I$_0$ between each RR Lyrae star 
is assumed to be only due to the variation in their distances and the relative distance, 
$\Delta$D, corresponding to each star is estimated. 
The error in the relative distance estimation of  
individual RR Lyrae star is basically the error in the extinction correction of I$_0$ mag. 
The average error in the extinction estimation, obtained from the RC stars, 
converts to a distance of $\sim$ 110 pc. 
The location of each RR Lyrae star in the cartesian coordinate system is then 
calculated  
using the transformation 
equations given in the previous section.

Using the dereddened I$_0$ magnitude of each RR Lyrae star, the dispersion 
in the surveyed region of the SMC can be found. The dispersion is a measure 
of the depth of the SMC. The data are binned in 0.5 degree in both x \& y axes 
to estimate dispersion. Thus the observed region is divided into bins with an area of 
0.25 square degrees. The number of stars in each bin 
have a range of 10 - 50. The bins with stars less than 10 are removed 
from our analysis. For each location we estimated 
the mean magnitude and the dispersion. 
The estimated dispersion have contributions 
from photometric errors, range in the metallicity of stars, intrinsic
variation in the luminosity due to evolutionary effects within the sample and 
the actual depth in the distribution of the stars \citep{c03}. 
We need to remove the contribution from the first three terms ($\sigma_{intrinsic}$) 
so that the value of the last term can be evaluated. \cite{c03}
estimated the value of $\sigma_{intrinsic}$ as 0.1 mag for their sample of 100 RRLS 
in the LMC. 
\cite{s06} used a larger sample of RRLS in the LMC from OGLE II catalog and estimated the value 
of $\sigma_{intrinsic}$ as 0.15 mag. Both these values are estimated using the globular 
cluster data in the observed field. In the case of the SMC, there is only one globular cluster 
and it is not in our observed field. So we used the intrinsic spread estimated for the 
RRLS in the LMC for the analysis of the RRLS in the SMC. 
Using both these values we estimated the corrected sigma values which 
correspond to only depth. The relation used is\\

$\sigma$$^2$$_{I_{0}mag}$ = $\sigma$$^2$$_{depth}$ + $\sigma$$^2$$_{intrinsic}$ 

\section{Reddening Map of the SMC}

One of the by-product of this study is the estimation of reddening towards the SMC. 
The observed shift of the peak (V$-$I) colour of the RC stars in the LMC, from the expected value 
was used by \cite{s05} to estimate the line of sight reddening map to the OGLE II 
regions of the LMC. Such a reddening map towards the SMC using the RC stars is obtained 
here. The intrinsic value of the (V$-$I) colour of the RC stars in the SMC is 
chosen as 0.89 mag to produce a median reddening equal to that measured by \cite{sc98} 
towards the SMC. Using this value the E(V$-$I) values are estimated as detailed in section 
2.1. The colour coded figure of the reddening in the SMC is presented in figure 4. The colour code 
is given in the plot. The average value of E(V$-$I) obtained towards the SMC is 0.053 $\pm$ 
0.017 mag. 
From the plot we can see that most of the regions in the SMC have E(V$-$I) less than 0.08 mag 
shown as cyan and blue points. The regions in the south western and north eastern sides of the center and the 
eastern wing regions have larger reddening compared to the 
other regions of the SMC. This reddening map estimated using the RC stars are used for dereddening the 
RC stars as well as the RRLS in this study. This reddening data will be made available electronically 
as an online table.

The previous estimates of the reddening towards the SMC are compared with our estimates. 
\cite{cc85} found a mean E(B$-$V) of 0.054 mag from the analysis of 48 Cepheids. This 
value can be converted into an E(V$-$I) of 0.0756 mag using the relation E(V$-$I) = 1.4 * E(B$-$V) \citep{s05}. 
\cite{u98} (using red clump stars) estimated 
a mean E(V$-$I) value of 0.1 mag. \cite{m95} and \cite{gm86} estimated 
a E(B$-$V) of around 0.09 mag which converts to an E(V-I) value of 0.126 mag. 
\cite{H11} provided the reddening maps towards the SMC  
from the study of the RC stars and RRLS. The intrinsic colour of the RC stars 
used by them is also 0.89 mag. The reddening map of the SMC estimated from the RC stars given in 
Figure 4 of \cite{H11} is very similar to our reddening map given in 
Figure 4. The mean value of E(V$-$I) estimated by them towards the SMC using the 
RC stars is 0.04 mag and that estimated from RRLS is 0.07 mag. The reddening maps of 
the RC stars and RRLS given in Figure 4 and Figure 11 of \cite{H11} are 
very similar. This fact supports and validates our usage of the reddening map obtained from 
the RC stars for dereddening the RRLS. 
The previous estimates other than that of \cite{H11} are higher than our estimates. 
This could be due to the choice of our intrinsic (V$-$I) colour of the RC stars 
in the SMC. But such a change only will shift the reddening value in all regions in a similar way. 
As we are only interested in the relative positions of the regions in the SMC such a change 
in the intrinsic colour of the RC stars is not going to affect our final results. 

In order to compare the adopted intrinsic colour of the RC stars in the SMC 
with theoretical values, we measured the peak colour of the RC stars from the 
synthetic CMD of the SMC given in \cite{GS01}. It turned out to 
be $\sim$ 0.89 mag. 
The table given in \cite{GS01} provide an RC (V$-$I)$_0$ colour  
for the LMC metallicity of z= 0.004 between 0.90 and 0.94 mag 
(age between 2-9 Gyr). For a lower metallicity of z = 0.001, their models 
lead to a colour between 0.8 mag and 0.84 mag (age between 2-9 Gyr). 
The mean metallicity found by Cole (1998) and Glatt et al (2008) 
for the SMC is between 0.002 and 0.003. This value supports our adopted 
value of (V$-$I)$_0$ colour of the RC stars in the SMC, which is bluer than 
that of the LMC and redder than the colour obtained for a lower metallicity system.  
 

\begin{figure}
\plotone{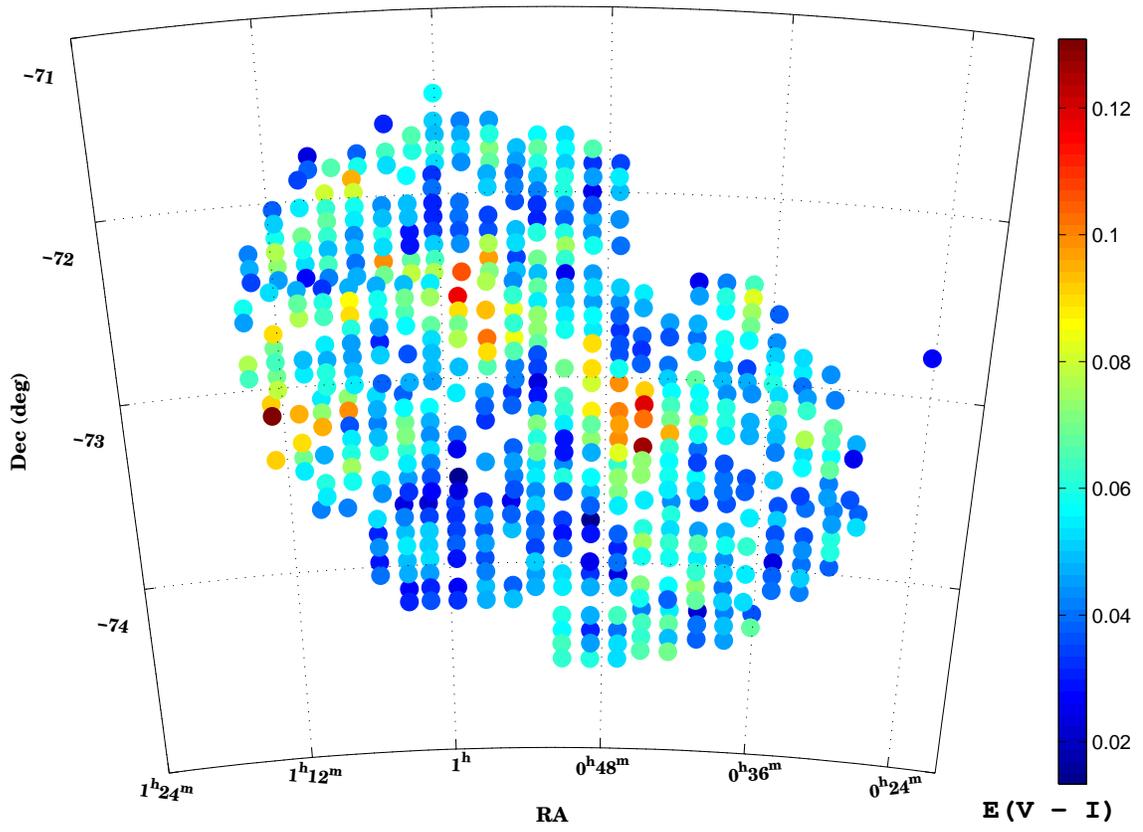}
\caption{A two dimensional plot of the reddening in the sub-regions of the SMC.}
\label{fig4} 
\end{figure}

\section{Results}

\subsection{Relative distances} 

The mean dereddened I$_0$ magnitude of the RC stars for the 
553 sub-regions of the SMC are estimated. 
The mean magnitudes of different sub-regions are shown as different colour points in the 
two dimensional plot of X vs Y in the figure 5. 
The colour code is given in the figure. The average dereddened magnitude, I$_0$ of 553 sub-regions 
is 18.43 $\pm$ 0.03 mag. The regions in the south (y$<0$) are fainter compared to the 
regions in the north. The brighter regions (I$_0$ $<$ 18.39) shown as blue points are located more 
in the northeastern side. This result indicates that 
the northeastern regions of the SMC are closer to us. 


\begin{figure}
\plotone{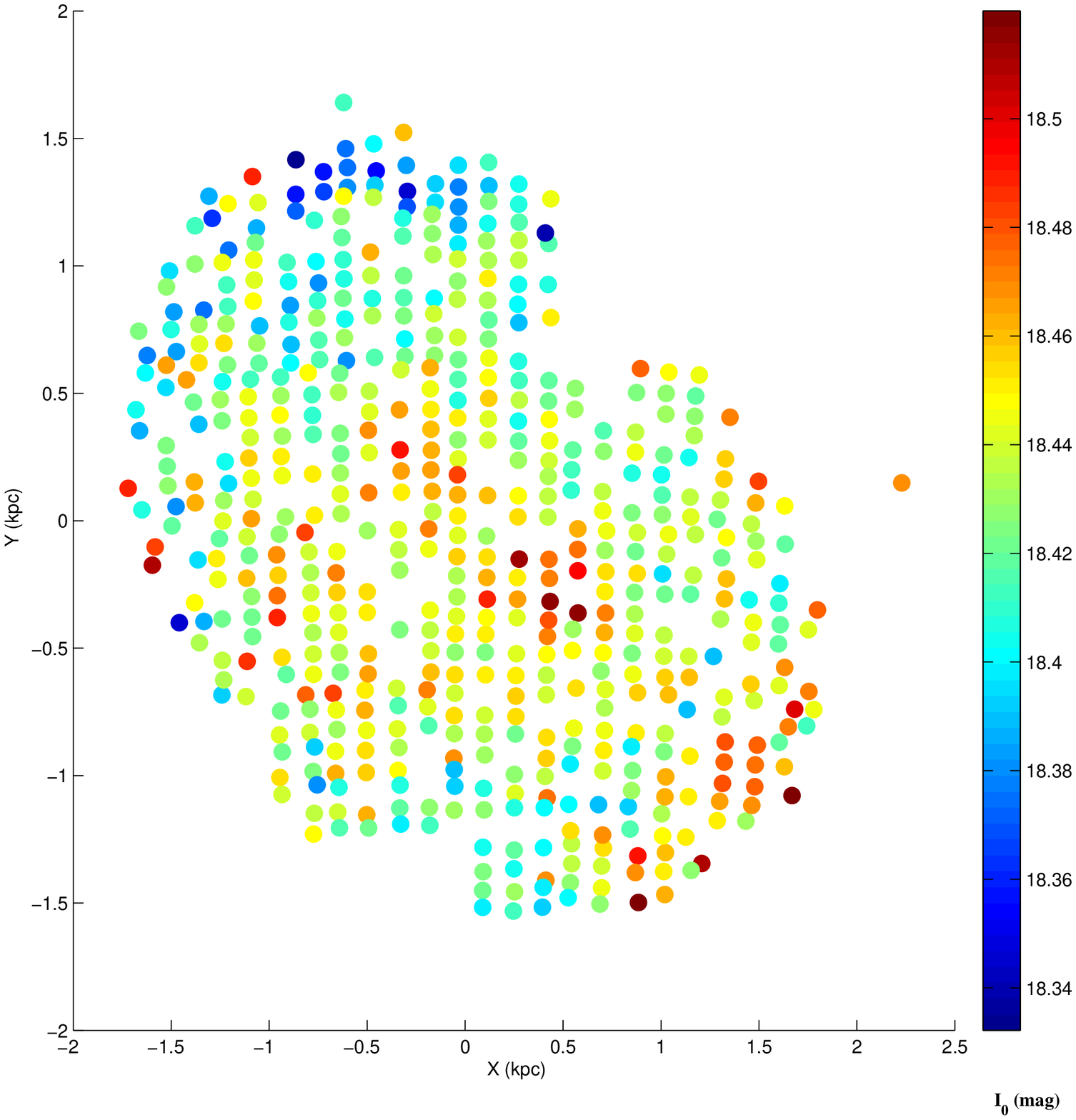}
\caption{A two dimensional plot of the mean magnitudes of the RC stars in the sub-regions of the SMC.}
\label{fig5} 
\end{figure}

The relative distance to each RR Lyrae star with respect to the mean distance to the SMC is 
estimated from the dereddened I$_0$ magnitude, assuming that the average distance to  
our sample of RRLS in the SMC is 60 kpc. 
The spatial distribution of RRLS in the XZ and YZ planes with over plotted density 
contours are shown in figure 6 and figure 7 respectively. 
The convention of the +ve and -ve Z axes are such that the +ve Z axis is towards us 
and -ve Z axis away from us. The distribution of RRLS 
in the SMC looks more or less symmetric and extends from -20 kpc to +20 kpc with 
respect to the mean. Most of the stars are located between $\pm$ 10 kpc. An extension 
to large distances can be seen at x $\sim$0. 
The outer density contours shown in figure 6 indicate that the eastern side of the 
SMC is on an average closer to us than the western side. We calculated the average 
I$_0$ magnitude of RRLS in the eastern end (x$<$-1.5, 245 stars) and in the western end 
(x$>$1.5, 221 stars). It was found that the eastern side is 0.034 mag brighter than the 
western side, which makes the eastern side on an average 950 pc closer 
to us than the western side. 
Similarly, the outer density contours shown in figure 7 suggest that the northern side 
is closer to us than the southern side. We calculated the average 
I$_0$ in the northern end (y$>$1.5, 133 stars) and in the southern end (y$<$-1.5,104 stars). It was found 
that the northern side is 0.019 mag brighter than the southern side, which makes the  
northern side on an average 520 pc closer to us than the southern side. 
Thus the outer density contours of figures 6 and 7 and the quantitative estimates mildly 
suggest that the north eastern part of the SMC is closer to us. 
This result is very similar to the result that obtained from the RC stars.   
As this effect is significant 
in outer regions and our study is limited to inner regions, we need more data 
in the outer region to securely say that the northeastern part of the SMC is closer 
to us. From the photometric 
study of stellar populations upto 11.1 kpc in the SMC, \cite{N11} found that the eastern 
part of the SMC is closer to us than the western side.
  
\begin{figure}
\plotone{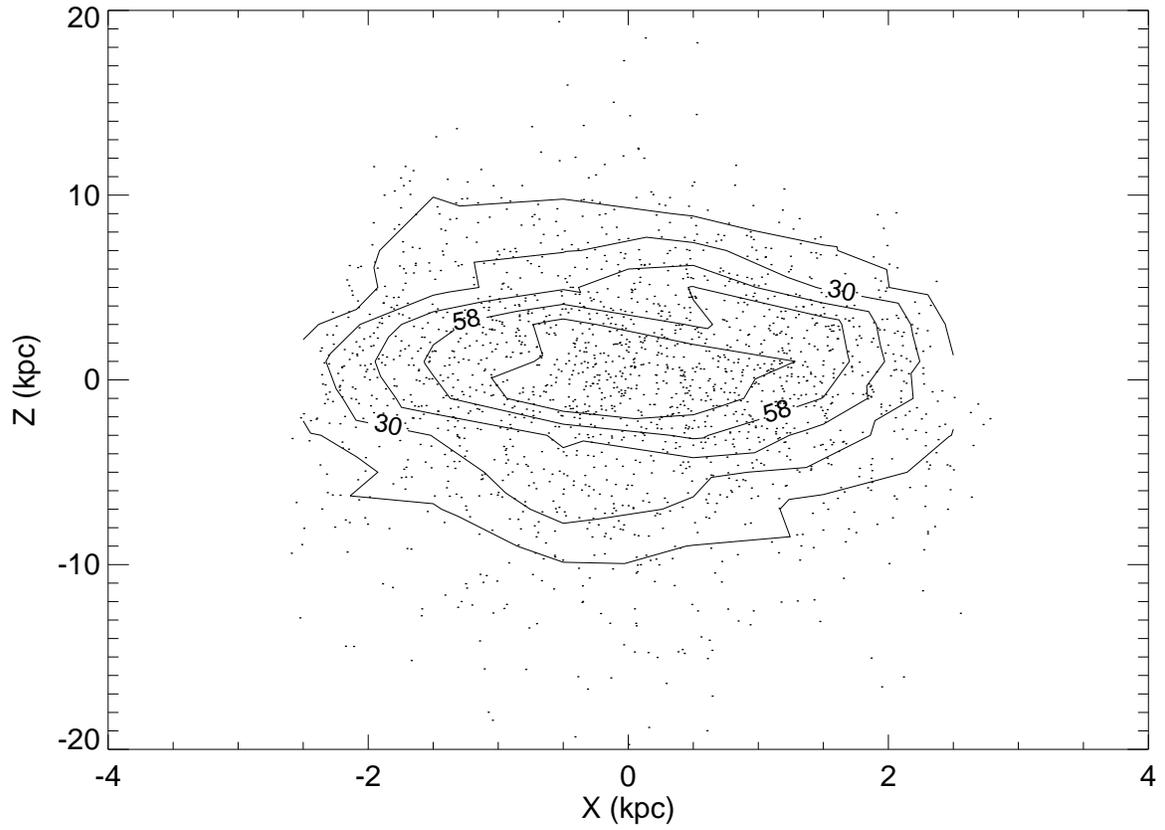}
\caption{Relative distances to each RR Lyrae star in the SMC with respect to the mean distance 
are plotted against X axis and are shown as black dots. The density contours are over plotted. The east 
is in the direction of decreasing X axis.}
\label{fig6} 
\end{figure}

\begin{figure}
\plotone{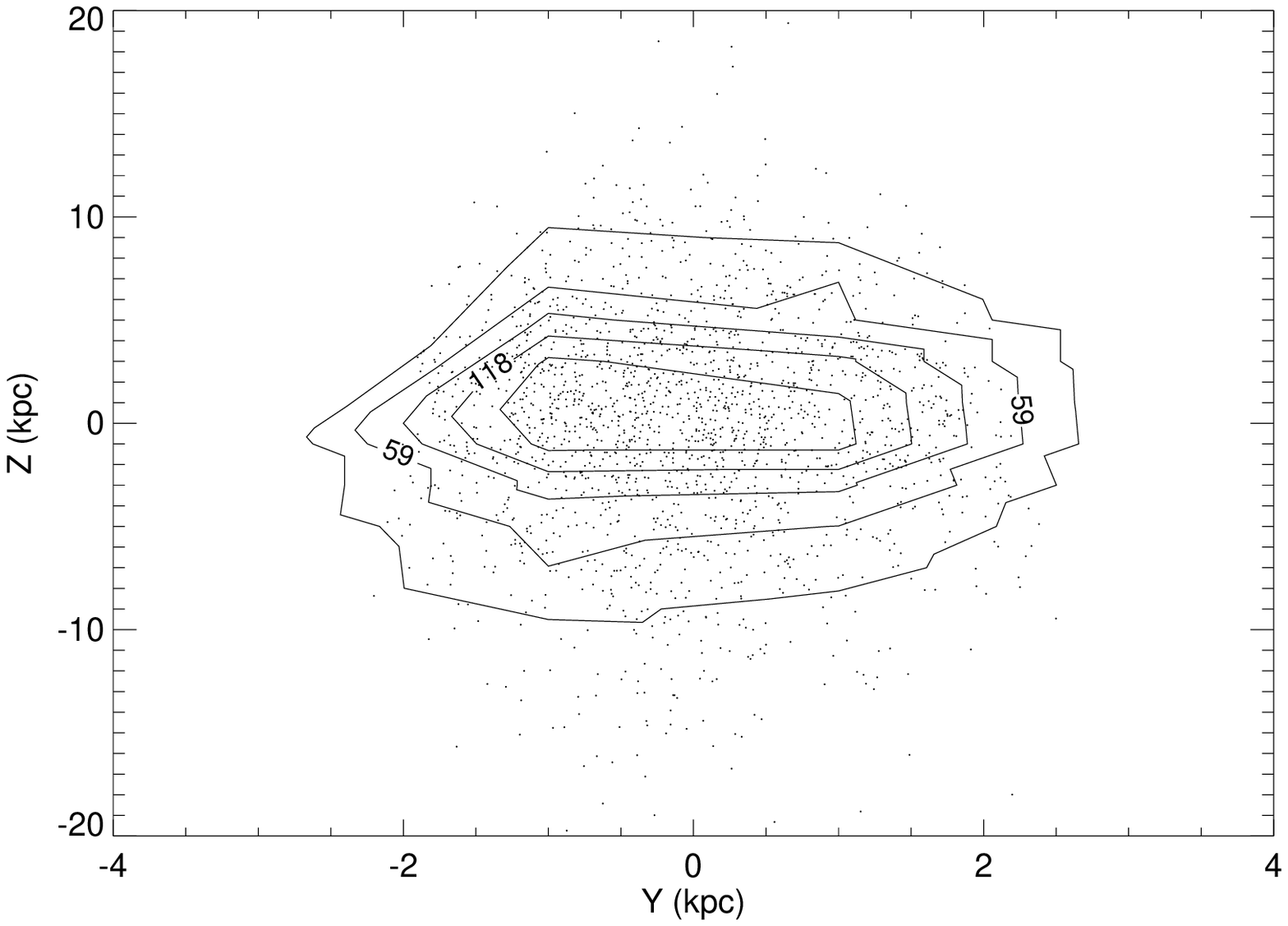}
\caption{Relative distances to each RR Lyrae star in the SMC with respect to the mean distance 
are plotted against Y axis and are shown as black dots. The density contours are over plotted. The north 
is in the direction of increasing Y axis.}
\label{fig7} 
\end{figure}

In the case of the RC stars as well as RRLS, the variation in the I$_0$ magnitude 
is assumed as only due to the difference in the distances. As discussed earlier in section 2, 
there can be contributions in their mean magnitude from the population effects. 
Thus the line of sight distance estimates suggest that either the RC star and RRLS in the 
north eastern part of the SMC  are  different from those found in the other regions of the SMC 
and/or the northeastern regions of the SMC are closer to us. The previous studies mentioned in section 2 suggest that there is no large variation in the 
age and metallcity among old stars in the inner SMC. Thus the contribution to the I$_0$ magnitude 
from the population effects across the SMC is expected to be minimum. The exact contribution 
can be understood only from the detailed spectroscopic studies of the old 
stellar populations.


 
\subsection{Line of sight depth of the SMC}

The dispersions in the magnitude and colour distributions of the RC stars are 
used to obtain the depth corresponding to the width (1-sigma) 
of the 553 sub-regions in the SMC. 
The width corresponding to depth is converted into depth in kpc using the distance 
modulus formula,\\
 
$\sigma_{depth}$ = 5log$_{10}$[(D$_0$ + d/2)/(D$_0$ - d/2)\\

where d is the line of sight depth in kpc.

A colour coded, two dimensional plot of depth is shown in figure 8. 
The colour code is explained in the plot. From the plot we can see that the depth 
distribution in the SMC is more or less uniform. The prominent feature 
in the plot is the enhanced depth (depth $>$ 8 kpc) seen near the central 
regions. An increased depth of around 6-8 kpc is seen near the north eastern 
regions also. The average depth obtained for the SMC observed region is 
4.57 $\pm$ 1.03 kpc. These results are matching well with the previous depth 
estimates by \cite{ss09}.

\begin{figure}
\plotone{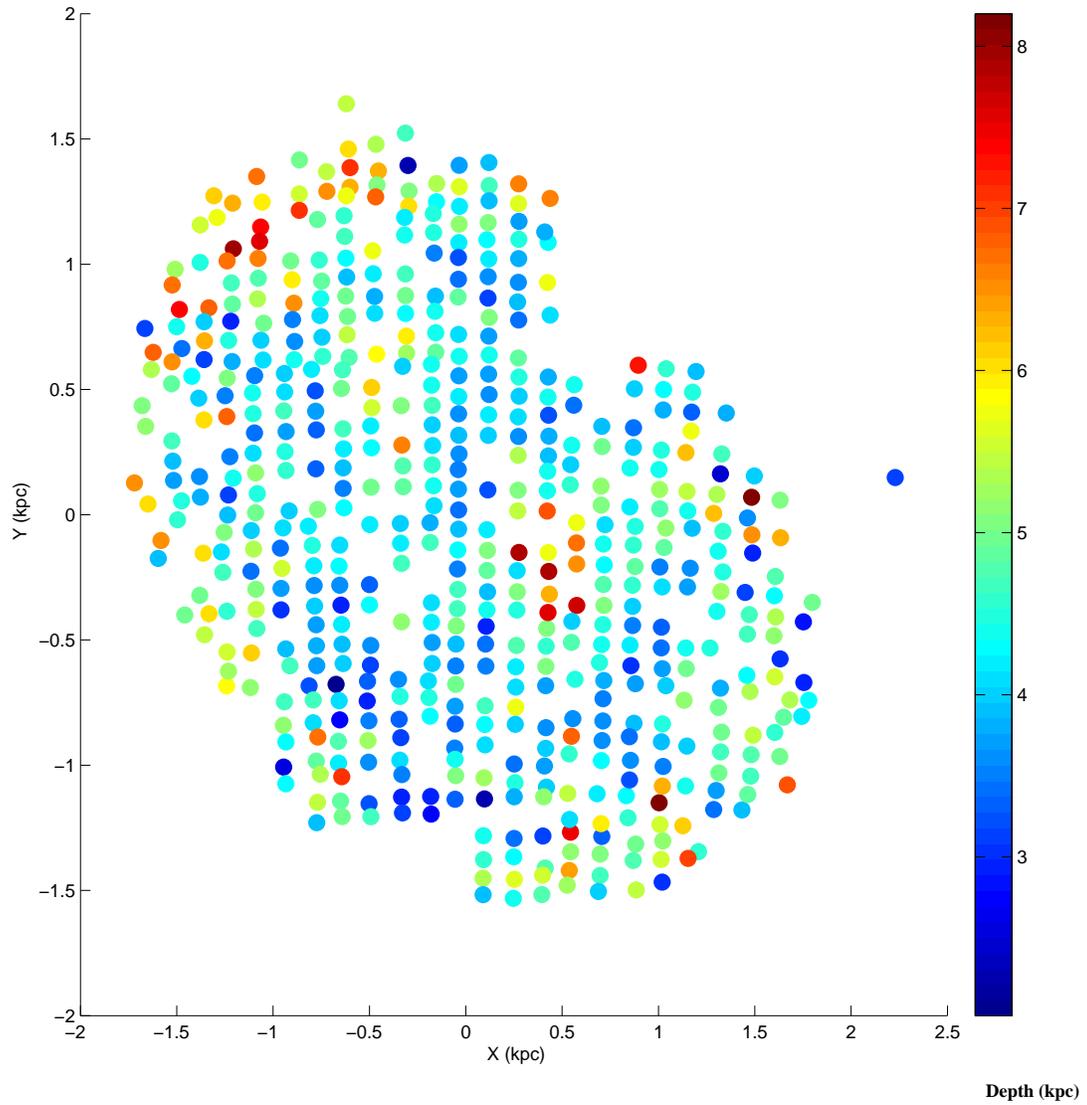}
\caption{A two dimensional plot of the depth in the sub-regions of the SMC 
obtained from the analysis of the RC stars.}
\label{fig8} 
\end{figure}

The dispersion in the mean magnitude of RRab stars is 
also used to estimate the depth (1-sigma) of the SMC. The 
dispersion in the mean magnitude for 70 sub-regions of the SMC 
is estimated and converted into depth in kpc. 
The average value of depth estimated when the intrinsic spread 
was taken as 0.1 mag is 4.07 $\pm$ 1.68 kpc and the average depth is 3.43 $\pm$ 1.82 kpc when 
0.15 mag was taken as the intrinsic spread. These estimates 
match with the depth estimate obtained from the RC stars within the error bars.

The depth estimated using the RC stars and RRLS are plotted together in figure 9 
against the X and Y axes. In both the lower panels, the depth values are 
plotted against the X-axis and in both the upper panels, they are plotted against 
the Y-axis.  
Basically the depth values correspond to the extent 
(front to back distance) over which these stars are distributed in the SMC.
The measured 1-sigma depth (front to back distance) is halved and plotted along 
the +ve and -ve depth-axis, assuming that the depth is symmetric with respect 
to the SMC. 
The red crosses correspond to the depth estimated using the RC stars and the black 
points correspond to the depth estimated using the RRLS. The error bars  
for each point are not plotted to avoid crowding.
In the left panels, the black dots correspond to the depth estimated 
using the RRLS when the $\sigma_{intrinsic}$ is taken as 0.1 mag. Similarly, in the 
right panels the black dots correspond to the depth estimated 
using the RRLS when the $\sigma_{intrinsic}$ is taken as 0.15 mag.
The larger depth near the 
center (x$\sim$0 and y$\sim$0) is seen for both the populations.
The 18 regions 
in the right panels and the 3 regions in the left panels which are shown as 
open circles at zero depth are those for which the depth of the RRLS 
became less than zero when the correction for $\sigma_{intrinsic}$ were applied. 
For a large number of regions the depth value becomes less than zero when 
$\sigma_{intrinsic}$ is taken as 0.15 mag. Thus, it appears to be more 
appropriate to take $\sigma_{intrinsic}$ as 0.1 mag for the RRLS in the SMC. 
The depth profiles of both the populations are similar, suggesting that both 
the RC stars and RRLS occupy a similar volume in the SMC.

\begin{figure}
\plotone{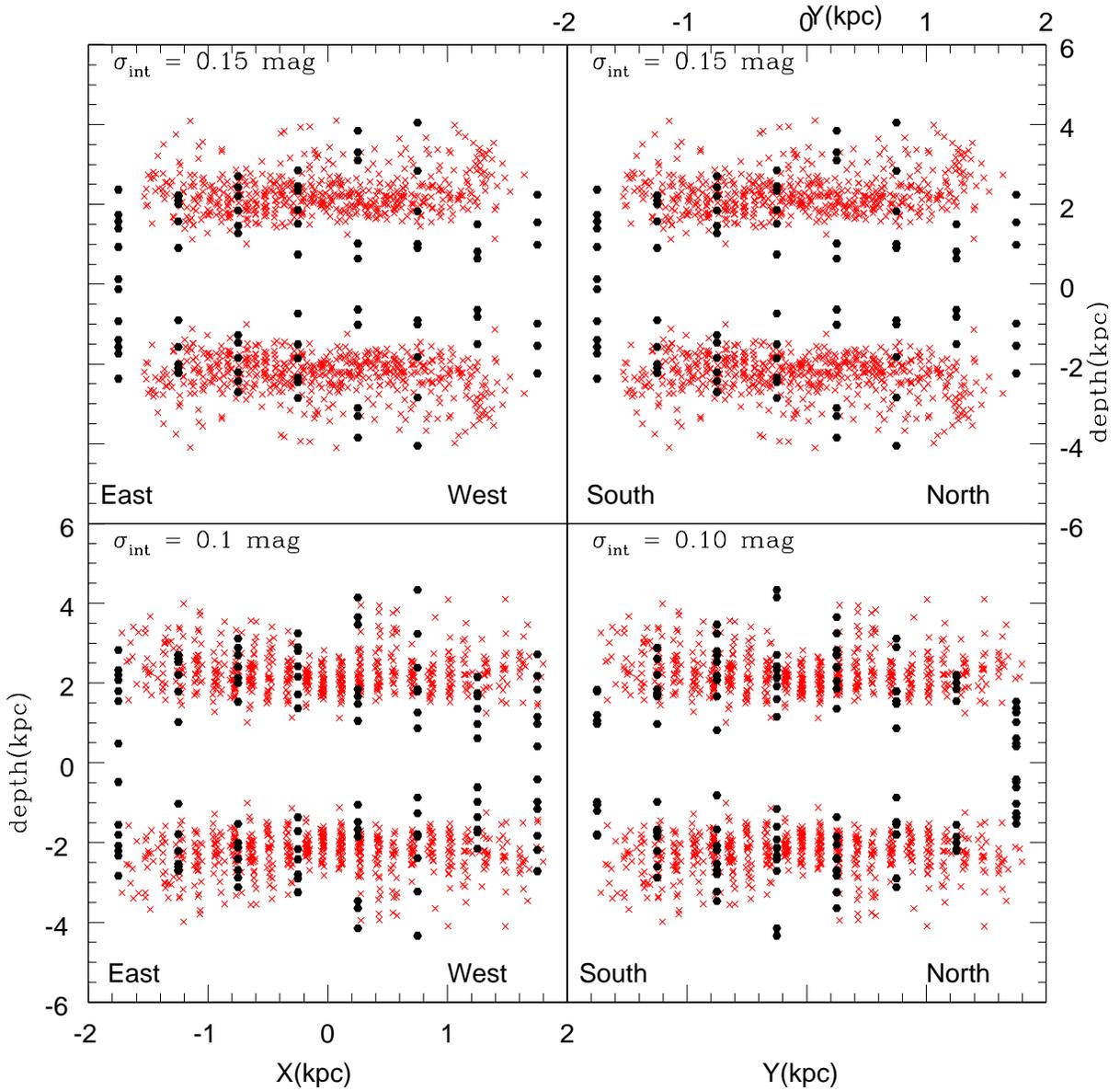}
\caption{The line of sight depth (1-sigma) of both the RC stars (red crosses) and the RRLS (black dots) 
are plotted against the X and Y axes. In the left lower and left upper panels, 
the RRLS depth estimates are done using $\sigma_{intrinsic}$ = 0.1 mag. In the right 
lower and left upper panels, the RRLS depth estimates are done 
using $\sigma_{intrinsic}$ = 0.15 mag.}
\label{fig9} 
\end{figure}

It is important to see how the dispersion in the distribution of the RRLS is 
related to the real RRLS distribution, obtained from the individual RRLS 
distances. This comparison will help us to estimate the actual depth from 
the dispersion of the RRLS in the SMC. We halved the 1-sigma depth 
(front to back distance estimated after correcting for 
$\sigma_{intrinsic}$ to be taken as 0.1 mag) 
obtained for each sub-region with respect to the mean distance and plotted 
in the -ve and +ve Z axis as open circles against the X-axis in the left 
lower panel of figure 10. Similarly, 2-sigma depth,3-sigma depth 
and 3.5-sigma depth are plotted in the lower right panel, upper-right panel and  
upper left panel of figure 10 respectively. The individual RRLS distances 
with respect to the mean SMC distance are shown as black dots in all the panels 
of figure 10. From the figure we can clearly see that the distribution 
of RRLS is atleast extended upto 3.5 sigma dispersion, if we take an 
intrinsic spread of 0.1 mag. This 3.5 sigma width ($\sigma_{dep}$ = 0.146 mag)
translates to a front to back distance of around 14.12 kpc. 
In figure 11, the depth estimated from RRLS after correcting for $\sigma_{intrinsic}$ 
= 0.15 mag are plotted over the individual RRLS distances. The left upper panel shows 
the 4 sigma depth plotted over the individual RRLS distances. From the figure we can 
clearly see that the distribution of RRLS is atleast extended to 4 sigma dispersion, if we 
take an intrinsic spread of 0.15 mag. This 4 sigma width 
($\sigma_{dep}$ = 0.124 mag) translates into a front to back distance of around 
13.7 kpc. Thus the RRLS in the SMC are distributed over a distance of $\sim$ 14 kpc 
along the line of sight.
In the case of the RC stars, we 
have estimated only the mean distances to the sub-regions and the depth of the 
sub-regions. So we cannot compare the real RC distribution with the width 
of the distribution to define the depth. Since the depth profiles of the 
RC stars and the RRLS are similar, we expect these two populations to occupy a similar 
volume in the SMC and the RC stars also to be distributed in a depth of 14 kpc.
The large depth suggests a spheroidal/ellipsoidal distribution for the above populations.

\begin{figure}
\plotone{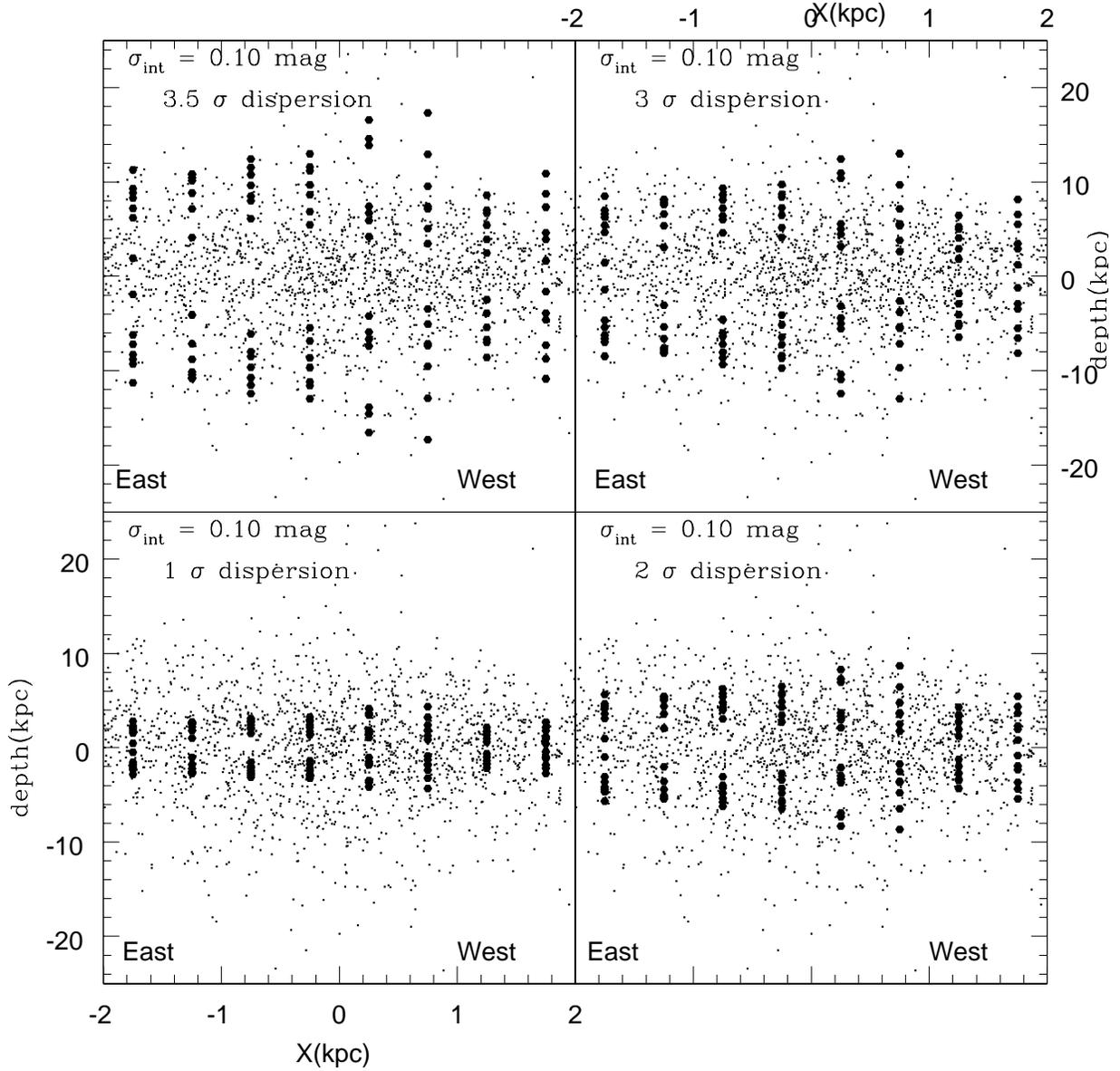}
\caption{The distance to each RRLS with respect to the mean distance to the SMC 
is plotted as black dots in all the panels against the X-axis. The open circles in 
all the panels from lower right to upper left in the counter clockwise direction 
 the 1-sigma, 2-sigma, 3-sigma \& 3.5-sigma depth halved with respect to the 
mean distance to each sub-region and plotted in the -ve and +ve Z-axis against 
the X axis. The depth is calculated using an intrinsic spread of 0.1 mag.}
\label{fig10} 

\end{figure}

\begin{figure}
\plotone{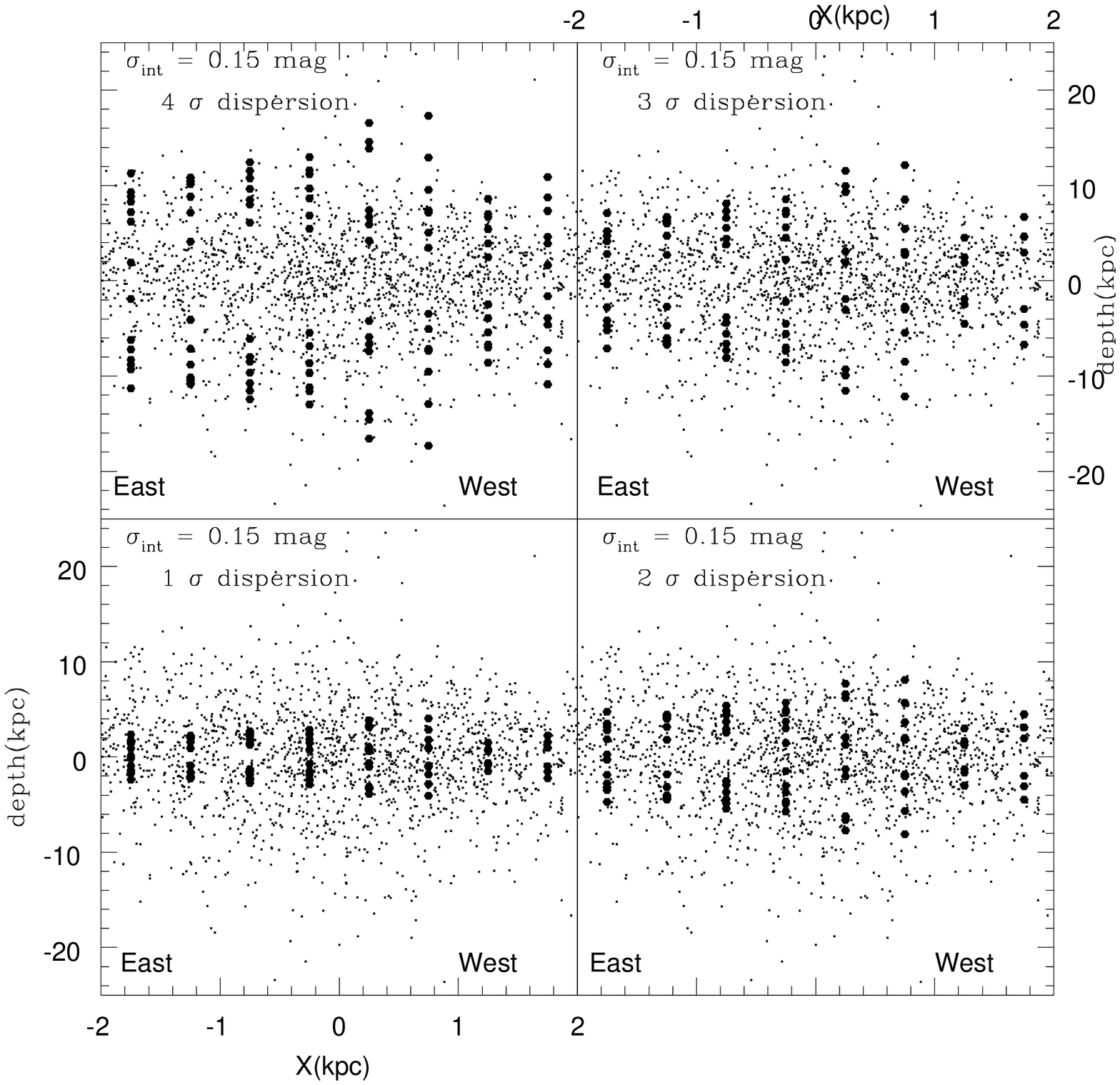}
\caption{The distance to each RRLS with respect to the mean distance to the SMC 
is plotted as black dots in all the panels against the X-axis. The open circles in 
all the panels from lower right to upper left in the counter clockwise direction 
 the 1-sigma, 2-sigma, 3-sigma \& 4-sigma depth halved with respect to the 
mean distance to each sub-region and plotted in the -ve and +ve Z-axis against 
the X axis. The depth is calculated using an intrinsic spread of 0.15 mag.}
\label{fig11} 

\end{figure}

\subsection{Density distributions of RC and RR Lyrae stars}
From the above section we found that both the RC stars and RRLS have 
similar line of sight depth. 
The density distributions of both these populations will give a clue 
about the system in the XY plane. The surface density distribution, and the 
radial density profile are studied to understand the structure of the SMC. The surface 
density is calculated by dividing the observed region into different sub-regions and 
obtaining the number of objects per unit area in each sub-region. The radial density profiles are 
obtained by finding the projected radial number density of objects in concentric rings 
around the centroid of the SMC. 
These profiles can be compared 
with the theoretical models. The two theoretical models which can be used for the comparison of 
spatial distribution of different stellar populations in the SMC are the exponential disk profile 
and the King's profile. The exponential disk profile is give by\\\\
f(r) = f$_{0d}$e${^{-r/h}}$\\\\
where f$_{0d}$ and the h represent the central density of the objects and the scale length, 
respectively and r is the distance from the centre of the distribution. 
The King's profile \citep{k62}, which is often used to describe the distribution of globular clusters, but also applies
to dwarf spheroidal galaxies, is given by\\\\
f(r) = f$_{0k}$\{[1+(r/a)$^2$]$^{-1/2}$ - [1+(r$_t$/a)$^2$]$^{-1/2}$\}$^2$\\\\
where r$_t$ and a are the tidal and core radii respectively, f$_{0K}$ is the central density of objects and
r is the distance from the centre. Both these profiles are used to fit the observed 
distribution of the RC stars in the SMC. 


\begin{figure}
\plotone{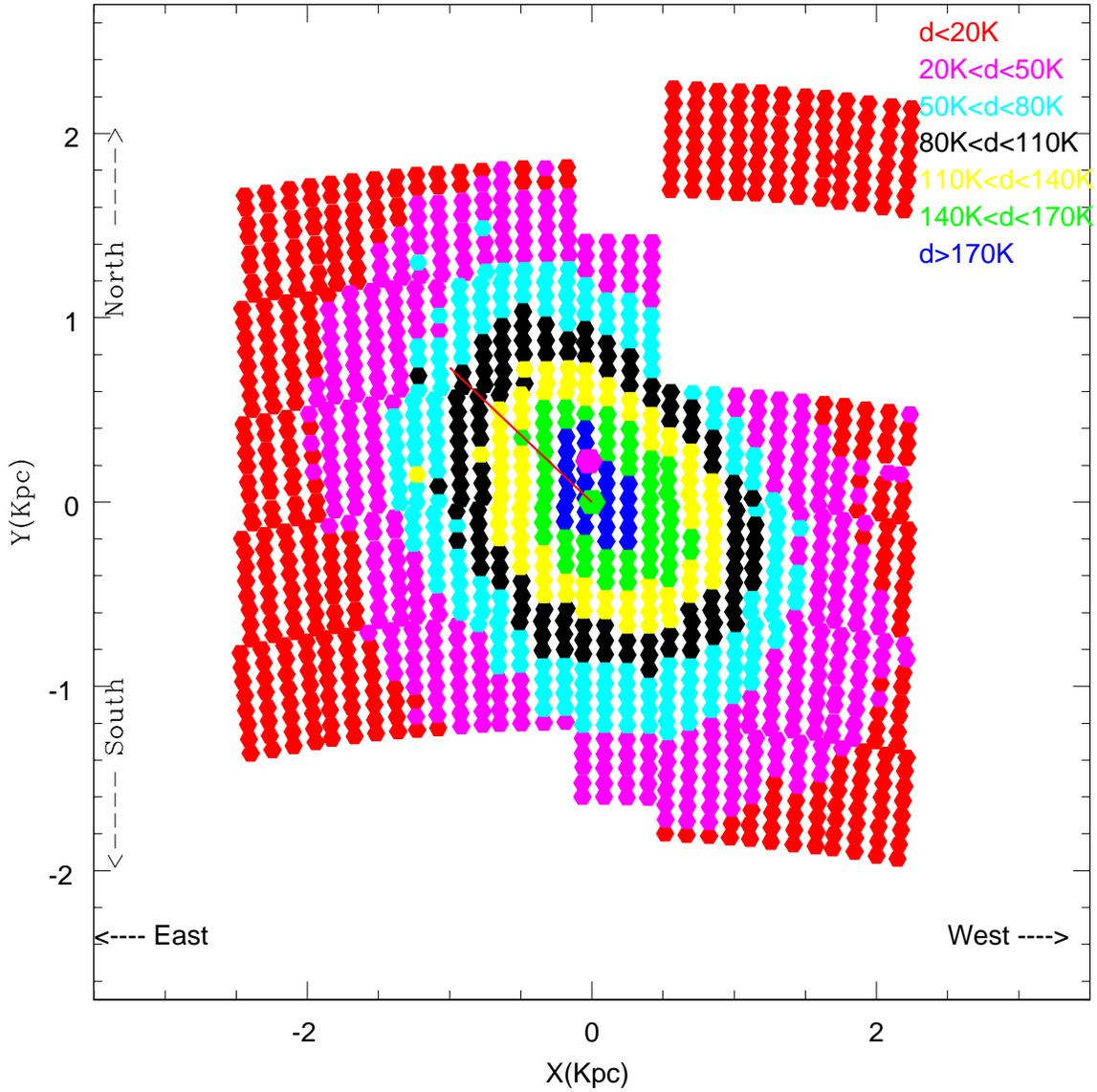}
\caption{A two dimensional plot of the number density,d of the RC stars 
(d in the units of 1000(K)/square kpc) in the SMC. The green and magenta 
hexagons represent the centroid of the sample and the optical center respectively. 
The red line shows the direction of elongation.}
\label{fig12} 
\end{figure}

The number of the RC stars, identified from the CMD, in each sub-region of the OGLE III 
region are estimated. We used all the 1280 sub-regions (each 
having an area of 32.6 square kpc) in the 
observed OGLE III region. We estimated the number density, number of the RC stars per 
unit area, for each sub-region. The number density distribution of the RC stars is shown in figure 12. 
The number density of RC stars in each sub-region ranges from 
7500/kpc$^2$ - 200,000/kpc$^2$. 
In this figure, discrete points represent each sub-region and the colour 
denotes the number density.  
The plot clearly shows the smooth RC distribution with an elongation in the 
north$-$east to south$-$ west (NE-SW) direction. 
The colour code is given in the figure, where d denotes the red clump number density. 
The position angle of the elongation of the RC distribution is around 54$^{\circ}$ for the eastern 
side. The plot also shows the shift in the RC density center,  $\alpha$ =  0$^h$ 52$^m$ 34$^s$.2,  
$\delta$ = -73$^{\circ}$ 2$^{'}$ 48$^{''}$ (shown as green hexagon in figure 12) 
from the optical center, $\alpha$ =  0$^h$ 52$^m$ 45$^s$,  $\delta$ = -72$^{\circ}$ 49$^{'}$ 43$^{''}$ 
(shown as magenta hexagon in figure 12) of the SMC. The center of the 
density distribution shown here is very similar to the centroid (given in section 2.1) of our RC sample 
in 553 sub-regions which are used for the estimation of dereddened magnitude and depth.

\begin{figure}
\plotone{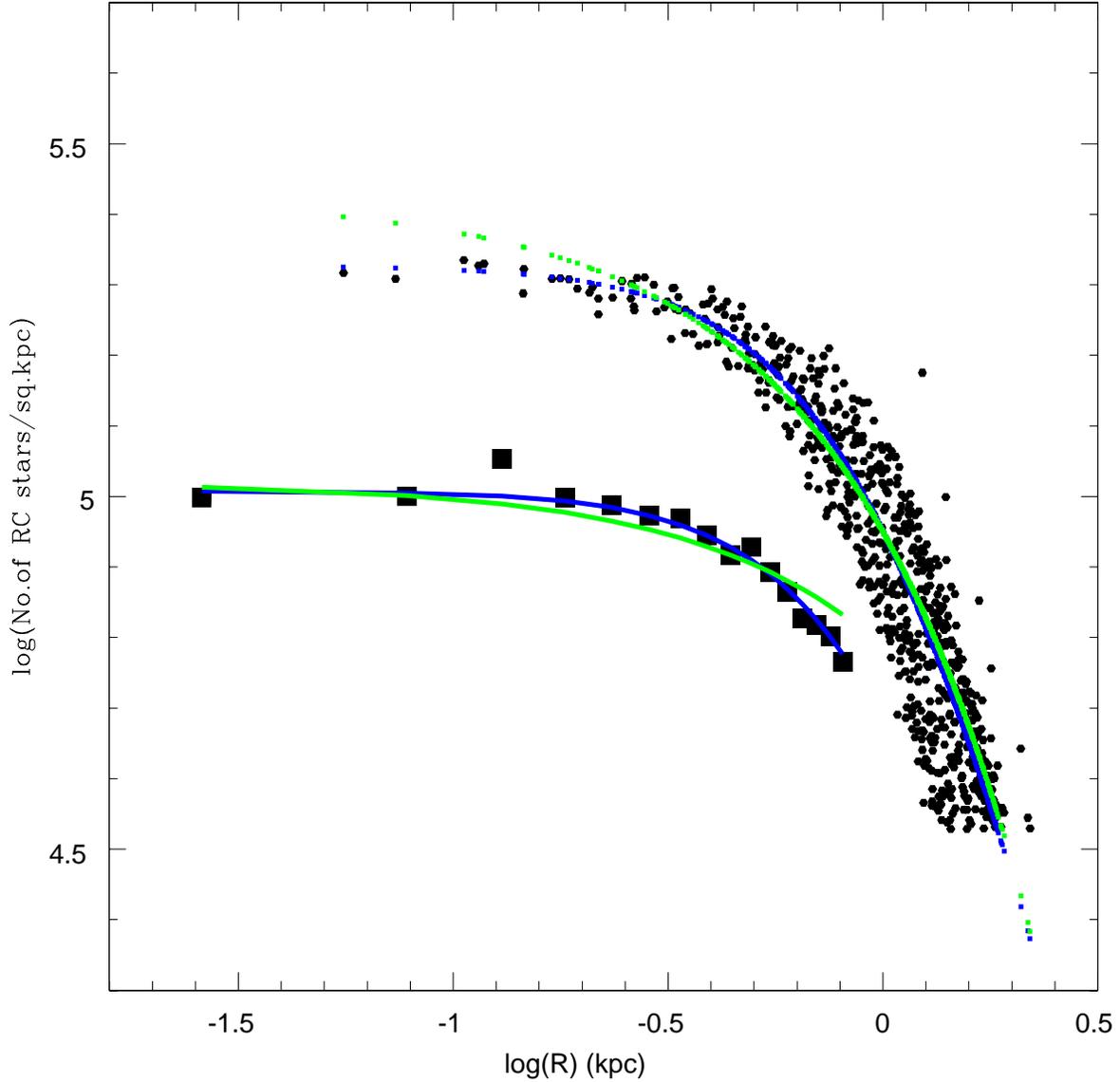}
\caption{Radial density distribution of the RC stars within the radius of the 0.8$^{\circ}$ in the SMC 
are shown as black squares. The best fitted King's profile and the exponential profile for the 
radial density distribution are shown as blue and green solid lines respectively. The surface density distribution 
of the RC stars for the whole observed region of the OGLE III are shown as black dots.  
The best fitted King's profile and the exponential profile for the surface density distribution are 
shown as blue and green dashed lines respectively. }
\label{fig13} 
\end{figure}

\begin{table*}
\centering
\caption{Parameters of the Exponential and King's profiles}
\label{Table:1}
\vspace{0.25cm}
\begin{tabular}{lrrrrrr}
 \hline
Data &  f$_{0d}$(no/sq.kpc) & h (kpc) & f$_{0k}$ (no/sq.kpc) & r$_{c}$(kpc) & r$_{t}$ 
(kpc)\\ 
\hline
Radial density distribution  & (104$\pm$0.4)x10$^3$ & 1.87$\pm$0.1 & 
(141$\pm$7)x10$^3$ & 1.08$\pm$0.02 & 7 $\pm$ 1\\
Surface density distribution &  (264$\pm$0.1)x10$^3$ & 0.91$\pm$0.1 
& (248$\pm$1)x10$^3$ & 0.9$\pm$0.01 & 12.04$\pm$ 0.01\\
\hline
\vspace{0.25cm}
\end{tabular}
\end{table*}

As the data coverage of the 
OGLE III is not symmetric with respect to the density center, the radial density distribution of the 
RC stars within a radius of 0.8$^{\circ}$ is obtained. The region within 0.8$^{\circ}$ radius from the 
density center are divided into 16 equal area (28.3 square arcmin) annuli. The number of RC stars in 
each annuli is obtained and it is divided by the area of the annuli to obtain the number density. 
Thus the radial density distribution of the RC stars within 0.8$^{\circ}$ radius from the density center 
is obtained and are shown as black squares in figure 13. The best fitted exponential disk profile and the 
King's profile are shown in the figure as green and blue solid lines respectively. We can see that the 
radial density profile of the RC stars in the SMC is marginally better described by the King's profile 
than by the exponential profile. We obtained the surface density profile of the RC stars in the whole 
observed area of the SMC. The number density of the RC stars in the 1280 sub-regions of the SMC are plotted 
against the radial distance of each sub-region from the density center in figure 13. The black points in the 
figure denotes the surface density of the RC stars in each sub-region. The best fit exponential disk 
profile and the King's profile are shown as green and blue dashed lines in figure 13. Here we can 
see that the surface density profile of the RC stars in the SMC is best described by the King's profile. 
The parameters obtained by fitting the radial density and surface density distributions by exponential 
and King's profiles are given in table 1. 
The estimated tidal radius of the SMC system is $\sim$ 7-12 kpc. 
Thus the RC stars in the SMC are distributed in a spherical/ellipsoidal volume, which indicates that the SMC can 
be approximated as a spheroidal/ellipsoidal galaxy. But in the surface density profile we can see a spread in the 
points for each radii, which indicates that the volume in which the RC stars are distributed is not 
exactly spherical. Again from the map of the surface density distribution shown in 
figure 12, we can see the elongation in the NE-SW direction. Thus the RC stars in the SMC are 
distributed in an ellipsoidal system.

Our sample of RRLS which are pulsating in the fundamental mode consists 1904 stars. 
The density center of our sample is $\alpha$ =  0$^h$ 53$^m$ 31$^s$,  $\delta$ = -72$^{\circ}$ 59$^{'}$ 15$^{''}$.7. 
The density center of our sample lies in between the two concentrations found by 
\cite{s10}.
To study the density profiles a large number of sample is required. 
The surface density map of the RRLS identified from the OGLE III photometric maps is 
shown in the lower panel of figure 7 in 
\cite{s10}. They identified two 
concentrations in the RRLS spatial distribution and found that the RRLS in the SMC form a roughly circular strucutre in the sky which 
indicates that the RRLS in the SMC are distributed in a spheroidal/ellipsoidal volume. 


\section{Axes ratio and orientation of the SMC ellipsoid}

We model the observed system of the SMC in which the RC stars and RRLS are distributed 
as a triaxial ellipsoid. 
The parameters of this ellipsoidal system, like the axes ratio and the orientation 
can be estimated using the inertia tensor analysis. The tensor analysis used here is 
similar to the methods used by \cite{ps09} and \cite{p06}, but 
with some modification. The tensor analysis used in the study is explained in 
appendix.

\subsection{RR Lyrae stars}
The method described in the appendix can be applied to our sample of the RRLS to estimate the parameters 
of the ellipsoidal component of the SMC. First we applied this method only to (x,y) system and 
found that the SMC RR Lyrae distribution is elongated with an axes ratio of 1:1.3 and the major axis 
has a position angle of 74$^{\circ}$ (NE-SW). 
We repeated the procedure to the (x,y,z) coordinates of the RRLS. The axes ratio 
obtained is 1:1.3:6.47 and the longest axis is inclined with the line of sight direction 
with an angle ($\it{i}$) of 0$^{\circ}$.4. The position angle of the 
projection of the ellipsoid ($\phi$) on the plane of the sky is given by 74$^{\circ}$.4.
The data of the RRLS used here contain the RRLS in the isolated north western OGLE III 
fields. As they are discrete points in the density distribution, we removed the RRLS 
in those fields and repeated the procedure. When we removed the RRLS in those 
fields, the density center changed to  $\alpha$ =  0$^h$ 54$^m$ 38$^s$.6,  
$\delta$ = -73$^{\circ}$ 4$^{'}$ 52$^{''}$.2. The axes ratio obtained is 1:1.57:7.71 
with  $\it{i}$ = 0$^{\circ}$.4 and $\phi$ = 66$^{\circ}$.0. Here we can see that the coverage of the 
data plays an important role in the estimation of the structural parameters 
of the ellipsoid. This may also suggest that there may be a variation in the inner and 
outer structures of the SMC. Nidever et al (2011) found that the inner SMC (R$<$3$^{\circ}$) is more elliptical 
than the outer component (3$^{\circ}$$<$R$<$7$^{\circ}$.5).   


In order to understand the radial variation of the inner structure 
we estimated the parameters using the data within 
different radii. At first, the analysis 
is done excluding the RRLS in the north western fields. The axes ratio, $\it{i}$ and $\phi$ 
of the RRLS distribution are obtained for the data within different radii, 
starting from 0.75$^{\circ}$. The values obtained are given in table 2. 
The values show that the $\it{i}$ is more or less constant with a value of around  0$^{\circ}$.5. 
But the axes ratio 
and $\phi$ have a range. The axes ratio has a range from 1:1.05:19.84 
to 1:1.57:7.7. The $\phi$ ranges from 60$^{\circ}$ to 78$^{\circ}$. The data within 0.75$^{\circ}$ of the 
density center is symmetric and the parameters obtained confirm that the RRLS distribution 
in the SMC is slightly elongated in the NE-SW 
direction. As the radius increases the data coverage is not symmetric and circular, 
and the axis ratio of x and y show an increasing trend. The surface density map of RRLS 
shown (in figure 7 of \citealt{s10}) does not show large elongation, rather 
the map looks nearly circular with a mild elongation in the NE-SW direction. 
So the elongation obtained for the RRLS distribution in the plane of the sky, 
at larger radius where the data coverage is not even, may not be real. 
We did a similar analysis to understand the 
radial variation including the fields in the north western regions. Here the concentric circles 
with different radii are centered on the density center (given in 
last paragraph) estimated including the RRLS in the north western fields. 
These  values are also given in table 2. From this also it is evident that 
there is a mild elongation in the NE-SW direction in the distribution of 
the RRLS in the SMC. Here we have to also keep in mind that the density center 
estimation of the SMC RRLS is not accurate as there are two concentrations found 
in the density distribution shown in figure 7 of \cite{s10}. 
The variation in the density center also modifies the structural parameters of the 
distribution. 

From the detailed analysis described in the last two paragraphs, it is clear that the 
quantitative estimates of the structural parameters of the SMC are very much dependent on the 
data coverage. Though there are differences in the values, we can say that the RR Lyrae distribution in the inner 
SMC is slightly elongated in the NE-SW direction. The longest axis Z', which is 
perpendicular to the X'Y' plane (a plane obtained by the counter clockwise rotation of the XY plane 
with an angle $\phi$ with respect to the Z axis) is aligned almost parallel to the line of sight, Z axis. 
The x, y and z values of the whole sample of the RRLS are plotted in figure 14 
to get a 3D visualization of the SMC structure. The figure clearly shows the 
elongation in the Z' $\sim$ Z axis. 

\begin{figure}
\plotone{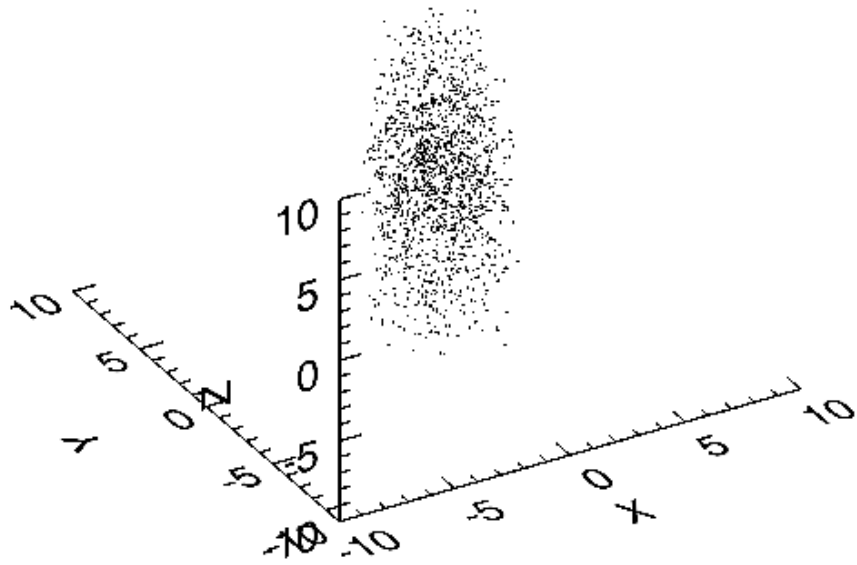}
\caption{The x,y and z values of the whole sample of the RRLS are plotted.}
\label{fig14} 
\end{figure}

\begin{table*}
\centering
\caption{Orientation measurements of the ellipsoidal component of 
the SMC estimated using the RR Lyrae stars}
\begin{tabular}{lrrrr}
\label{Table:2}
Excluding the three north western fields\\
Center  $\alpha$ =  0$^h$ 54$^m$ 38$^s$.6, $\delta$ = -73$^{\circ}$ 4$^{'}$ 52$^{''}$.2\\
\hline
 Data  &  No of RRLS & Axes ratio &  $\it{i}$ & $\phi$\\ 
\hline
 r $<$ 0$^{\circ}$.75 &  421 & 1:1.05:19.84 &  0$^{\circ}$.4 & 78$^{\circ}$.83\\
 r $<$ 1$^{\circ}$.00 &  676 & 1:1.03:14.59 &  0$^{\circ}$.4 & 72$^{\circ}$.20\\
 r $<$ 1$^{\circ}$.25 &  924 & 1:1.04:11.35 &  0$^{\circ}$.2 & 75$^{\circ}$.68\\
 r $<$ 1$^{\circ}$.50 & 1187 & 1:1.10:9.43  &  0$^{\circ}$.1 & 63$^{\circ}$.41\\
 r $<$ 1$^{\circ}$.75 & 1407 & 1:1.23:8.66  &  0$^{\circ}$.1 & 61$^{\circ}$.68\\
 r $<$ 2$^{\circ}$.00 & 1563 & 1:1.34:8.21  &  0$^{\circ}$.1 & 66$^{\circ}$.00\\
 r $<$ 2$^{\circ}$.25 & 1711 & 1:1.47:7.87  &  0$^{\circ}$.3 & 67$^{\circ}$.62\\
 r $<$ 2$^{\circ}$.50 & 1780 & 1:1.54:7.73  &  0$^{\circ}$.4 & 67$^{\circ}$.48\\
 r $<$ 2$^{\circ}$.75 & 1798 & 1:1.57:7.71  &  0$^{\circ}$.4 & 66$^{\circ}$.30\\
 r $<$ 3$^{\circ}$.00 & 1803 & 1:1.57:7.71  &  0$^{\circ}$.4 & 65$^{\circ}$.96\\
 
\hline\\
Including the north western fields\\
Center  $\alpha$ =  0$^h$ 53$^m$ 31$^s$,  $\delta$ = -72$^{\circ}$ 59$^{'}$ 15$^{''}$.7\\
\hline
Data  &  No of RRLS & Axes ratio &  $\it{i}$ & $\phi$\\ 
\hline
 r $<$ 0$^{\circ}$.75 &  428 & 1:1.07:20.01 &  0$^{\circ}$.5 & 48$^{\circ}$.84\\
 r $<$ 1$^{\circ}$.00 &  671 & 1:1.03:14.01 &  0$^{\circ}$.4 & 64$^{\circ}$.88\\
 r $<$ 1$^{\circ}$.25 &  927 & 1:1.03:11.19 &  0$^{\circ}$.4 & 67$^{\circ}$.84\\
 r $<$ 1$^{\circ}$.50 & 1177 & 1:1.13:9.59 &  0$^{\circ}$.2 & 59$^{\circ}$.54\\
 r $<$ 1$^{\circ}$.75 & 1399 & 1:1.23:8.70  &  0$^{\circ}$.2 & 60$^{\circ}$.10\\
 r $<$ 2$^{\circ}$.00 & 1568 & 1:1.30:8.00  &  0$^{\circ}$.1 & 64$^{\circ}$.87\\
 r $<$ 2$^{\circ}$.25 & 1755 & 1:1.34:7.22  &  0$^{\circ}$.3 & 68$^{\circ}$.78\\
 r $<$ 2$^{\circ}$.50 & 1845 & 1:1.36:6.89  &  0$^{\circ}$.4 & 70$^{\circ}$.80\\
 r $<$ 2$^{\circ}$.75 & 1892 & 1:1.34:6.57  &  0$^{\circ}$.4 & 73$^{\circ}$.70\\
 r $<$ 3$^{\circ}$.00 & 1904 & 1:1.33:6.47  &  0$^{\circ}$.3 & 74$^{\circ}$.40\\
\end{tabular}
\end{table*}

\begin{table*}
\centering
\caption{Orientation measurements of the ellipsoidal component of the SMC using the RC stars}
\label{Table:3}
\vspace{0.25cm}
\begin{tabular}{lrrrr}
 \hline
 Data &   Axes ratio &  $\it{i}$ & $\phi$\\ 
\hline
RC stars within 1-sigma depth & 1:1.49:3.08 &  4$^{\circ}$.22 & 55$^{\circ}$.7\\
RC stars within 2-sigma depth & 1:1.48:5.26 &  1$^{\circ}$.23 & 55$^{\circ}$.4\\
RC stars within 3-sigma depth & 1:1.48:7.10 &  0$^{\circ}$.68 & 55$^{\circ}$.6\\
RC stars within 3.5-sigma depth & 1:1.48:7.88 & 0$^{\circ}$.58 & 55.$^{\circ}$.5\\
RC stars within 4-sigma depth & 1:1.48:8.58 & 0$^{\circ}$.49 & 55.$^{\circ}$.4\\
\hline
\vspace{0.25cm}
\end{tabular}
\end{table*}

\subsection{Red Clump stars}
From the previous sections we found that the RC stars and RRLS occupy a similar volume of 
the SMC. Hence we can apply the method of inertia tensor to the RC stars for the 
estimation of the parameters of the ellipsoidal component of the SMC. First we apply this 
method to the (x,y) coordinates of the 553 sub-regions of the SMC. Here each (x,y) pair 
represents the coordinates of a sub-region. We weighted it using the number of 
the RC stars identified in that region to get the axes ratio. We find 
that the RC distribution in the SMC is elongated with an axes ratio of 1:1.48 
and the position angle of the major axis is 55$^{\circ}$.3. Now we applied the same procedure 
to the (x,y,z) system of the RC stars. In the case of RRLS, we took the distance of 
the individual RR Lyrae star with respect to the mean distance as the z coordinate. 
But in the case of RC stars we only have the mean magnitudes corresponding 
to the mean distances to the sub-regions. The mean magnitude of a region in the 
ellipsoidal or nearly spheroidal system is the average of the magnitude of symmetrically 
distributed RC stars. In order to get the real RC distribution we did the 
following. After subtracting the average extinction, we obtained the magnitude distribution 
of the RC stars for each sub-region. Using the method explained earlier in 
section 2.1, we estimated the peak and the width of the magnitude 
distribution for each sub-region and eventually the width corresponding to depth. 
The 
depth (front to back distance) of the RC stars in a region is the measure of the extend 
up to which the RC stars in that region is distributed. Most of the RC 
stars in a sub-region are at the mean distance obtained for that sub-region 
and the remaining are distributed within the depth of the region. We assume the 
distribution to be symmetric with respect to the mean distance to that sub-region. 
Initially we took the I$_0$ values of the bins within 1-sigma width and converted 
them into z distances. As we have the number of stars in each bin we weighted and 
applied the inertia tensor analysis to the x,y and z coordinates. 

\begin{table*}
\centering
\caption {Orientation measurements of the ellipsoidal component of the SMC using the RC stars and the RRLS in the same region.
 Excluding the north western fields and the edges of the data} 
\label{Table:4}
\vspace{0.25cm}
\begin{tabular}{lrrrr}
 \hline
 Data &   Axes ratio &  $\it{i}$ & $\phi$\\ 
\hline
1654 RRLS  & 1:1.46:8.04 & 0$^{\circ}$.50 & 58$^{\circ}$.3\\
RC stars within 3.5-sigma depth & 1:1.48:7.88 &  0$^{\circ}$.58 & 55$^{\circ}$.5\\
\hline
\vspace{0.25cm}
\end{tabular}

\end{table*}

Earlier we found that the RC and RR Lyrae depth distributions are similar. Also, 
from figures 10 and 11 we can see that the RR Lyrae distribution extends up to 3-4 sigma 
depth. Hence we can take the z component to extend upto higher sigma levels (2-sigma, 3-sigma, 
3.5-sigma, 4-sigma)
 to understand the real RC distribution in the SMC. 
We applied the inertia tensor analysis using z components which extend up to 
2-sigma, 3-sigma, 3.5-sigma \& 4-sigma. The parameters obtained are given in table 3. From the table 
we can see that the orientation measurements of the ellipsoidal component of the 
SMC using RC stars, where the z component extend up to 3.5 sigma is similar to the 
values obtained for the RR Lyrae distribution excluding the north western fields. 
As we removed the regions with RC stars less than 400, 
the north western fields are not included in the analysis. The axes ratio and the 
angle $\it{i}$ are matching well with the ellipsoidal parameters of the SMC, 
estimated using the distribution of the RRLS. The position angle, $\phi$ 
estimated using the RC stars and RRLS are slightly different. 

Based on the analysis described above we can say that the RC stars in the SMC are 
distributed in an ellipsoidal system with an axes ratio of 1:1.48:7.88. The position angle, $\phi$ 
is 55$^{\circ}$.5. The longest axis Z' which is perpendicular 
to the X$'$Y$'$ plane (obtained by the counter clockwise rotation of the XY plane with an angle 55$^{\circ}$.5 
with respect to the Z axis) is inclined with the Z axis with an angle of 0$^{\circ}$.58$^{\circ}$. The inclination 
of the Z$'$ axis with the line of sight is very small and we can assume the longest axis to be almost 
along the line of sight.

\subsection{Comparison of the structural parameters obtained from both the populations}

The quantities estimated in this analysis strongly depend on the data coverage. In order to 
compare the axes ratio and the orientation measurements 
obtained from both the populations, it is important to take the sample of both 
the population of stars from the same region. In the case of the RC stars 
regions which include the north western fields and also regions in the edges of the data set 
are omitted based on the selection criteria.  
Such a truncation of the data is not done in the case of the RRLS. Only the RRLS which are possible 
Galactic objects are removed from the analysis. These stars are not concentrated on a particular region but are 
scattered. If we remove the RRLS in the north western fields and the edges of the data, then the 
sample of RRLS cover the nearly the same region as the RC stars sample. We used the RRLS within the box of size size -1.8$-$1.8 
in both the X and Y axes. Thus the RRLS in the edges of the data set and in the north western fields 
are nearly removed. 
Then we estimated the parameters using this sample of the RRLS and the values are 
given in table 4. 
From the table we can see that when the sample of both the populations are taken from similar regions 
of the SMC the estimated parameters match well.

\subsection{The actual structure of the SMC}
\begin{table}
\centering
\caption{Estimated structural parameters with equal extent in all the three axes}
\label{Table:1}
\vspace{0.25cm}
\begin{tabular}{lrrrr}
 \hline
 Data &   Number of RRLS & Axes ratio &  $\phi$ (degrees) & $\it{i}$ (degrees)\\ 
\hline
 $\rho$ $<$ 2.0   &   344    & 1:1.17:1.28 &  67.5  &  4.2\\
 $\rho$ $<$ 2.5   &   540    & 1:1.24:1.39 &  69.5  &  3.3\\
 $\rho$ $<$ 3.0   &   730   &  1:1.33:1.61 &  70.2 &   2.6\\
\hline
\vspace{0.25cm}
\end{tabular}
\end{table}
We find the Z' $\sim$ Z axis is the longest axis in the analysis of both 
the RRLS and the RC stars. The reason for such a result is mainly the coverage 
of the SMC. The studied data covers only the central regions 
thus restricting the coverage in the XY plane. On the other hand the full Z direction 
is sampled. Hence the present view of the SMC based on our axes ratio is like 
viewing only the central part of a sphere along the Z axis.
Such a perspective will give an elongated Z axis. This is clear from figure 14. 
In order to test whether the z'$\sim$z axis is due to the viewing perspective, 
we have sampled the data similarly along the three axes. The extend of the sample along all the three axes can be made equal by taking the 
data only within a spherical radii ($\rho$) of 2, 2.5 and 3 degrees from the center. 
We took the RRLS stars within these different radii and estimated the structural 
parameters. The estimated parameters are given in table 5. An elongation in 
the NE-SW axis is seen suggesting that the elongation of the RRLS in the NE-SW is a real feature. 
The decrease in the relative length of the z' axis is expected 
as the selection mentioned above removes relatively more number of 
stars along the z' direction than in the x' and y' directions. 
From table 5 we can see that the inclination of the 
longest axis with the line of sight axis decreases from inner to outer radii.  
The finite XY coverage restricts further analysis of this trend. The 
very small value of $\it{i}$ (given in tables 2, and 4) estimated from 
the analysis of all the RRLS, including those at larger z distances from the center, 
indicates that the decreasing trend of $\it{i}$ is continued to outer radii also. 
In table 5 we can also see an increase in $\phi$ and the relative length of y' axis 
from inner to outer region. From section 6.1 and from table 2 we can understand that 
these two quantities are dependent on the choice of the centroid of the system. So we 
cannot definitely say anything more about this trend. Another important result to be noticed from table 5 is that 
even though the z' axis is the longest of the three, the values of the longest and 
the second longest (y') axes are comparable. This suggests that when the coverage 
along all the three axes becomes comparable, the structure of the SMC is 
spheroidal or slightly ellipsoidal.

The above analysis indicates that the XY extent of the SMC upto which 
stellar populations are studied/detected plays an important role in understanding  
the actual structure of the SMC and hence the estimation of the structural parameters.  
The tidal radius estimates give the radial extent up to which the stellar populations are bound to the SMC or up to which they are 
expected to be found. \cite{gn96} estimated the tidal radius of the 
SMC to be $\sim$ 5 kpc.  Later \cite{s04} suggested the 
tidal radius of the SMC to be 4-9 kpc. The estimations obtained 
in our study from the radial density and surface density profiles of the RC stars, 
given in table 1, suggest the tidal radius to be 7-12 kpc.   
There are many recent studies which found old and intermediate stellar populations in the 
outer regions.  
\cite{NG07} found that up to 6.5 kpc from the SMC center, the galaxy is composed of both, 
intermediate-age and old population, without an extended halo. \cite{r10} estimated the SMC edge 
in the eastern direction to be around 6 kpc from the survey of red giant stars in 10 
fields in the SMC. A very recent photometric survey of the stellar periphery of the SMC 
by \cite{N11} found the presence of old and intermediate age populations 
at least up to 9 kpc. The above mentioned observational studies and tidal radius estimates 
suggest that the full extent of the SMC in the XY plane (2*9 = 18 kpc) is of the order of  
the front to back distance estimated (14 kpc) along the Z axis. 
We found in earlier sections that the inner SMC is slightly elongated in the NE-SW direction. 
Thus we suggest, that the actual structure of the SMC is spheroidal or slightly 
ellipsoidal. Better estimates of 
the structural parameters can be obtained with a data set of larger sky coverage. 
In the following section, we combine previous studies and
suggest an evolutionary model for the SMC. 



\section{Discussion}
\cite{bk08b} suggested that the SMC may be formed due to a 
dwarf-dwarf merger, resulting in a stellar spheroidal component and a gaseous disk 
component, happened before the formation of the MW-LMC-SMC system. They also suggest 
that since stellar populations formed before the merger event should have dynamically hot
kinematics, the youngest age of stellar populations
that show both spheroidal distributions and no or little 
rotation can correspond to the epoch when the merger occurred. 
As the merger event should evoke a large scale star formation event in the 
galaxy, the epoch of merger can also be traced from the enhancements seen in the 
star formation history of the SMC.

The most comprehensive study of the star formation history of the SMC is presented 
by \cite{hz04}. They derived the global star formation history of the 
SMC. They found that there was a significant epoch of star formation up to
8.4 Gyr ago when $~$50$\%$ of the stars were formed, followed by
a long quiescent period in the range 3 Gyr $<$ age $<$ 8.4 Gyr,
and a more or less continuous period of star formation starting
3 Gyr ago and extending to the present. They also found three
peaks in the SFR, at 2$-$3 Gyr, at 400 Myr, and 60 Myr ago. Their CMDs do not go 
deep enough to derive the full star formation history from the information 
on the main sequence. Obtaining CMDs reaching the oldest main sequence  
turnoff is essential in order to properly constrain the intermediate-age and 
old population \citep{g05}. As we study the intermediate age and old stellar 
populations in the SMC, here we compare the star formation studies done based on the 
CMDs which reach the oldest main sequence turnoffs. These studies go deeper and hence 
have smaller field of views. \cite{d01}, \cite{cv07}, 
and \cite{n09} have done such studies and obtained the 
star formation history of the SMC.

\cite{d01} found a broad peak of star formation between 5 and 8 Gyr ago. 
\cite{cv07} found two main episodes of star formation , 
at 300$-$400 Myr and between 3$-$6 Gyr. They also found that 
the star formation rate was low until $\approx$ 6 Gyr ago, when only a few
stars were formed. \cite{n09} found star formation enhancements 
at two intermediate ages, a conspicuous one peaked at 4$-$5 Gyr old in all 
fields and a less significant one peaked at 1.5$-$2.5 Gyr old in all fields. 
The enhancement at old ages, with the peak at 10 Gyr old is seen in all fields. But 
in the western fields, this old enhancement is split into two at 8 Gyr
old and at 12 Gyr old. Their farthest field is at 4.5 kpc. All these 
studies show an enhancement in the star formation in the SMC around 4$-$5 Gyrs ago. 

If the 4$-$5 Gyr global star formation in the SMC is assumed to be due to 
a dwarf-dwarf merger, then according to the merger model, the merged galaxy 
will settle down within 1-3 Gyr \citep{l08}. This scenario will then demand a 
spheroidal component well mixed with stars older than 2 Gyr.
Our results find that the old population (RR Lyrae stars, age $>$ 9 Gyr)
and the intermediate age component (Red clump stars, age = 2$-$9 Gyr) seem 
to occupy similar volume, suggesting that stars older than 2 Gyr are in a well 
mixed ellipsoid. 

\cite{t09} proposed that the evidence of a major merger event in the SMC 
is imprinted in the age-metallicity relation as a dip in [Fe/H]. They predicted 
that the major merger with a mass ratio of 1:1 to 1:4 occurred at $\sim$ 7.5 Gyr 
ago in the SMC. Based on this model they could reproduce the abundance distribution 
function of the field stars in the SMC. As they could not correlate a peak 
in the star formation history of the SMC at the epoch of the merger they suggested 
that the major merger which occurred in the SMC at 7.5 Gyr ago did not trigger 
a major star burst due to some physical reasons but proceeded with a moderate star 
formation. \cite{n09} did not find a dip in the age metallicity relation of the SMC,  
which \cite{t09} claim to detect and suggested as the imprint of a major merger at 7.5 Gyr ago. 
Again if this merger event in the SMC at 7.5 Gyr old is the reason for the 
kinematical and morphological differences of the young and old stars then we cannot 
expect stars younger than 4-5 Gyr in the spheroidal/ellipsoidal distribution. 
As we see the RC stars which have an age range of 2-9 Gyr in the 
spheroidal/ellipsoidal distribution we propose a merger event which started 
around 4-5 Gyr ago as the reason for the observed distribution of stars in the SMC. 
 

A wider and larger future photometric (OGLE IV) and spectroscopic surveys which 
cover both the inner and outer regions of the SMC will help to understand the 
complete structure and kinematics of the SMC.

\section{Conclusions}

$\bullet$ The dereddened I${_0}$ magnitude of the RC stars and RRLS are used 
to determine the relative positions of the regions in the SMC with respect to the 
mean distance and it suggest that either the population of the RC stars and RRLS in the 
north eastern regions are different and/or the north eastern part of the SMC is closer to us.\\

$\bullet$ The line of sight depth of the SMC estimated using the RC stars and 
the RRLS is found to be $\sim$ 14 kpc. The depth profiles of both 
the population look similar indicating that these two populations are 
located in the similar volume of the SMC.\\ 

$\bullet$ The surface density distribution and the radial density profile of the RC 
stars suggest that they are distributed in a nearly spheroidal system. 
The tidal radius estimated for the SMC system is $\sim$ 7-12 kpc.
An elongation 
from NE -SW is also seen in the surface density map of the RC stars in the SMC.\\ 

$\bullet$ The observed SMC is approximated as a triaxial ellipsoid and the structural 
parameters, like the axes ratio, inclination, $\it{i}$ of the longest axis with the line of 
sight and the position angle, $\phi$ of the projection of the ellipsoid 
on the plane of the sky are estimated using the inertia tensor analysis. 
From the analysis of the RC stars and RRLS in the same region of the SMC 
the parameters estimated turned out to be very similar. The estimated 
parameters are very much dependent on the data coverage.\\ 


$\bullet$ The study of data only within concentric spheres of radii 2, 2.5 and 
3 degrees from the center, where the extent in all 
the three axes becomes equal, shows that the relative lengths of the z' and y' 
axes are comparable. 
We estimated an axes ratio of 1:1.33:1.61 with a 2$^\circ$.6 inclination of the 
longest axis with the line of sight from the analysis of RRLS within 3 degrees in the 
X, Y and Z axes. The position angle of the projection of the ellipsoid on the sky 
obtained is 70$^\circ$.2.
Our tidal radius estimates and various observational studies of the 
outer regions suggest that the full extent of the SMC in the XY plane is 
similar to the front to the back distance estimated along the line of sight. 
These results suggest that the actual structure of the SMC is spheroidal or slightly 
ellipsoidal.\\

$\bullet$ We propose that the SMC experienced a merger with another dwarf galaxy 
at about 4$-$5 Gyr ago, and the merger process was completed in another 2-3 Gyr. 
This resulted in a spheroidal distribution comprising of stars older than 2 Gyr.\\

\acknowledgements
Smitha Subramanian acknowledges the financial support provided by Council of 
Scientific and Industrial Research (CSIR) , India through SRF grant, 
09/890(0002)/2007-EMR-I. The authors thank the OGLE team for making the 
data available in public. Thanks to Indu.G for her help 
in plotting figures. The authors thank the anonymous referee for the constructive 
suggestions which improved the manuscript.

\section{Appendix}
\begin{center}
{\bf Inertia tensor analysis}
\end{center}
Moment of inertia of a body characterizes the mass distribution within the body. For any rotating 
three dimensional system, we can compute the moment of inertia about the axis of rotation, 
which passes through the origin of a local reference (XYZ) frame, using the inertia tensor. 
The origin of the system is the center of mass of the body. Consider a system made up of 
i number of particles, each particle with a mass m. For each particle the (x,y and z) coordinates 
with respect to the center of mass is known. Then, the moment of inertia tensor, I of the system is 
given by\\

  I = $\left(
                \begin{array}{ccc}
                I_{xx} & I_{xy} & I_{xz}\\
                I_{yx} & I_{yy} & I_{yz}\\
                I_{zx} & I_{zy} & I_{zz}\\
                \end{array}
                \right)$\\
where\\
 I$_{xx}$ = $\Sigma{_i}$ m$_{i}$(y$_{i}$$^2$ + z$_{i}$$^2$)\\
 I$_{yy}$ = $\Sigma{_i}$ m$_{i}$(x$_{i}$$^2$ + z$_{i}$$^2$)\\
 I$_{zz}$ = $\Sigma{_i}$ m$_{i}$(x$_{i}$$^2$ + y$_{i}$$^2$)\\
 I$_{xy}$ = I$_{yx}$ = $\Sigma{_i}$ m$_{i}$(x$_{i}$ + y$_{i}$)\\
 I$_{yz}$ = I$_{zy}$ = $\Sigma{_i}$ m$_{i}$(y$_{i}$ + z$_{i}$)\\
 I$_{xz}$ = I$_{zx}$ = $\Sigma{_i}$ m$_{i}$(x$_{i}$ + z$_{i}$)\\

The components, I$_{xx}$,I$_{yy}$ and I$_{zz}$ are called the moments of inertia while I$_{xy}$, 
I$_{yx}$, I$_{xz}$, I$_{zx}$, I$_{yz}$ and I$_{zy}$ are the products of inertia. These components 
given above are basically specific to the local reference frame and reflect the mass 
distribution within the system in relation to the local reference frame. If we align the axes of 
the local reference frame in such a way that the mass of the system is evenly distributed around 
the axes then the product of inertia terms vanish. This 
would mean the transformation 
of the local reference frame (XYZ) to a system (X'Y'Z') . In the new frame of reference, inertia 
tensor, I' is given by\\

I' =  $\left(\begin{array}{ccc}
             I_{x'x'} & 0 & 0\\
             0 & I_{y'y'} & 0\\
             0 & 0 & I_{z'z'}\\
             \end{array}
             \right)$\\

The terms I$_{x'x'}$, I$_{y'y'}$ and I$_{z'z'}$ are the non zero diagonal terms of the inertia 
tensor in the new reference frame (X'Y'Z') and are called the principal moments of inertia of the 
body. The three axes in the new reference frame are called the principal axes of the body.

To determine the principal axes of the system we have to diagonalize the inertia tensor I, which 
is obtained with respect to the local reference frame. The diagonalization of the inertia tensor 
provides three eigen values ($\lambda_1$, $\lambda_2$ and $\lambda_3$) which correspond 
to the moments of inertia ( I$_{x'x'}$, I$_{y'y'}$ and I$_{z'z'}$) about the principal axes. 
For each eigen value value we can compute the corresponding eigen vector. 
The eigen vectors corresponding to each eigenvalues are given as  
e$_{\lambda_1}$ = e$_{11}$ i' + e$_{12}$ j' + e$_{13}$ k', 
e$_{\lambda_2}$ = e$_{21}$ i' + e$_{22}$ j' + e$_{23}$ k' \&
e$_{\lambda_3}$ = e$_{31}$ i' + e$_{32}$ j' + e$_{33}$ k' 
where i', j' and k' are unit vectors along the X', Y' and Z' axes respectively.
The transformation of the XYZ system to X'Y'Z' can be obtained 
using the transformation matrix, T which is made up of the 9 components of the 3 eigen vectors. 
The transformation matrix, T is given by

         T =  $\left(
                \begin{array}{ccc}
                e_{11} & e_{21} & e_{31}\\
                e_{12} & e_{22} & e_{32}\\
                e_{13} & e_{23} & e_{33}\\
                \end{array}
                \right)$\\

From the eigen values and the transformation matrix, the axes ratio and the orientation of the 
characteristic ellipsoid that best describes the spatial distribution of the particles in the 
system can be obtained. As the moment of inertia of a system characterizes the resistance 
of the system to rotation, the component of moment of inertia along the major axis of the 
system will be least and maximum along the minor axis. Thus the three eigen values which 
represent the moments of inertia of the three axes (such that I$_{x'x'}$$>$I$_{y'y'}$$>$I$_{z'z'}$) 
of the ellipsoid can be written as
\\
I$_{x'x'}$ = M(a$^2$ + b$^2$) \\
I$_{y'y'}$ = M(a$^2$ + c$^2$) \\
I$_{z'z'}$ = M(b$^2$ + c$^2$) \\
where a,b \& c are the semi-axes of the ellipsoid such that a$>$b$>$c and M the total mass of the 
system. Using the above relations we can estimate the axes ratio of the ellipsoid which best 
describes the spatial distribution of the particles in the system. The transformation matrix, T 
describes the spatial directions or the orientation of the ellipsoid with respect to the 
local reference frame.


\begin{thebibliography}{52}
\expandafter\ifx\csname natexlab\endcsname\relax\def\natexlab#1{#1}\fi

\bibitem[{{Bekki} \& {Chiba}(2008)}]{bk08b}
{Bekki}, K. \& {Chiba}, M. 2008, \apjl, 679, L89

\bibitem[{{Borissova} {et~al.}(2009){Borissova}, {Rejkuba}, {Minniti},
  {Catelan}, \& {Ivanov}}]{b09}
{Borissova}, J., {Rejkuba}, M., {Minniti}, D., {Catelan}, M., \& {Ivanov},
  V.~D. 2009, \aap, 502, 505

\bibitem[{{Caldwell} \& {Coulson}(1985)}]{cc85}
{Caldwell}, J.~A.~R. \& {Coulson}, I.~M. 1985, \mnras, 212, 879

\bibitem[{{Caldwell} \& {Coulson}(1986)}]{cc86}
---. 1986, \mnras, 218, 223

\bibitem[{{Chiosi} \& {Vallenari}(2007)}]{cv07}
{Chiosi}, E. \& {Vallenari}, A. 2007, \aap, 466, 165

\bibitem[{{Cioni} {et~al.}(2000){Cioni}, {Habing}, \& {Israel}}]{c00}
{Cioni}, M., {Habing}, H.~J., \& {Israel}, F.~P. 2000, \aap, 358, L9

\bibitem[Cioni et al (2006)]{C06SMC} Cioni, M.-R.L., Girardi.L., Marigo.P., Habing, 
H.J., 2006, A\&A, 452, 195

\bibitem[{{Clementini} {et~al.}(2003){Clementini}, {Gratton}, {Bragaglia},
  {Carretta}, {Di Fabrizio}, \& {Maio}}]{c03}
{Clementini}, G., {Gratton}, R., {Bragaglia}, A., {Carretta}, E., {Di
  Fabrizio}, L., \& {Maio}, M. 2003, \aj, 125, 1309

\bibitem[Cole(1998)]{C98}Cole, A. A. 1998, ApJ, 500, L137

\bibitem[{{Crowl} {et~al.}(2001){Crowl}, {Sarajedini}, {Piatti}, {Geisler},
  {Bica}, {Clari{\'a}}, \& {Santos}}]{c01}
{Crowl}, H.~H., {Sarajedini}, A., {Piatti}, A.~E., {Geisler}, D., {Bica}, E.,
  {Clari{\'a}}, J.~J., \& {Santos}, Jr., J.~F.~C. 2001, \aj, 122, 220

\bibitem[{{De Propris} {et~al.}(2010){De Propris}, {Rich}, {Mallery}, \&
  {Howard}}]{r10}
{De Propris}, R., {Rich}, R.~M., {Mallery}, R.~C., \& {Howard}, C.~D. 2010,
  \apjl, 714, L249

\bibitem[{{Dolphin} {et~al.}(2001){Dolphin}, {Walker}, {Hodge}, {Mateo},
  {Olszewski}, {Schommer}, \& {Suntzeff}}]{d01}
{Dolphin}, A.~E., {Walker}, A.~R., {Hodge}, P.~W., {Mateo}, M., {Olszewski},
  E.~W., {Schommer}, R.~A., \& {Suntzeff}, N.~B. 2001, \apj, 562, 303

\bibitem[{{Dopita} {et~al.}(1985){Dopita}, {Lawrence}, {Ford}, \&
  {Webster}}]{d85}
{Dopita}, M.~A., {Lawrence}, C.~J., {Ford}, H.~C., \& {Webster}, B.~L. 1985,
  \apj, 296, 390

\bibitem[{{Evans} \& {Howarth}(2008)}]{eh08}
{Evans}, C.~J. \& {Howarth}, I.~D. 2008, \mnras, 386, 826

\bibitem[{{Gallart} {et~al.}(2005){Gallart}, {Zoccali}, \& {Aparicio}}]{g05}
{Gallart}, C., {Zoccali}, M., \& {Aparicio}, A. 2005, \araa, 43, 387

\bibitem[{{Gardiner} \& {Hawkins}(1991)}]{gh91}
{Gardiner}, L.~T. \& {Hawkins}, M.~R.~S. 1991, \mnras, 251, 174

\bibitem[{{Gardiner} \& {Noguchi}(1996)}]{gn96}
{Gardiner}, L.~T. \& {Noguchi}, M. 1996, \mnras, 278, 191

\bibitem[Giraradi \& Salaris (2001)]{GS01} Girardi, L. \& Salaris, M. 2001, MNRAS, 323, 109

\bibitem[{{Graham}(1975)}]{g75}
{Graham}, J.~A. 1975, \pasp, 87, 641

\bibitem[{{Grieve} \& {Madore}(1986)}]{gm86}
{Grieve}, G.~R. \& {Madore}, B.~F. 1986, \apjs, 62, 427

\bibitem[{{Groenewegen}(2000)}]{g00}
{Groenewegen}, M.~A.~T. 2000, \aap, 363, 901

\bibitem[{{Harris} \& {Zaritsky}(2004)}]{hz04}
{Harris}, J. \& {Zaritsky}, D. 2004, \aj, 127, 1531

\bibitem[{{Harris} \& {Zaritsky}(2006)}]{hz06}
---. 2006, \aj, 131, 2514

\bibitem[{{Hatzidimitriou} {et~al.}(1993){Hatzidimitriou}, {Cannon}, \&
  {Hawkins}}]{h93}
{Hatzidimitriou}, D., {Cannon}, R.~D., \& {Hawkins}, M.~R.~S. 1993, \mnras,
  261, 873

\bibitem[{{Hatzidimitriou} {et~al.}(1997){Hatzidimitriou}, {Croke}, {Morgan},
  \& {Cannon}}]{h97}
{Hatzidimitriou}, D., {Croke}, B.~F., {Morgan}, D.~H., \& {Cannon}, R.~D. 1997,
  \aaps, 122, 507

\bibitem[Haschke et al (2011)]{H11}Haschke,R., Grebel.E.K., \& Duffau,S. 2011, AJ, 141
\bibitem[{{King}(1962)}]{k62}
{King}, I. 1962, \aj, 67, 471

\bibitem[{{Lotz} {et~al.}(2008){Lotz}, {Jonsson}, {Cox}, \& {Primack}}]{l08}
{Lotz}, J.~M., {Jonsson}, P., {Cox}, T.~J., \& {Primack}, J.~R. 2008, \mnras,
  391, 1137

\bibitem[{{Maragoudaki} {et~al.}(2001){Maragoudaki}, {Kontizas}, {Morgan},
  {Kontizas}, {Dapergolas}, \& {Livanou}}]{m01}
{Maragoudaki}, F., {Kontizas}, M., {Morgan}, D.~H., {Kontizas}, E.,
  {Dapergolas}, A., \& {Livanou}, E. 2001, \aap, 379, 864

\bibitem[{{Massey} {et~al.}(1995){Massey}, {Lang}, {Degioia-Eastwood}, \&
  {Garmany}}]{m95}
{Massey}, P., {Lang}, C.~C., {Degioia-Eastwood}, K., \& {Garmany}, C.~D. 1995,
  \apj, 438, 188

\bibitem[Nidever et al (2011)]{N11}Nidever, D.L., Majewski, S.R., Munoz, R.R., et al, 
2011, ApJ, 733, L10 
                                                        ̃


\bibitem[{{No{\"e}l} {et~al.}(2009){No{\"e}l}, {Aparicio}, {Gallart},
  {Hidalgo}, {Costa}, \& {M{\'e}ndez}}]{n09}
{No{\"e}l}, N.~E.~D., {Aparicio}, A., {Gallart}, C., {Hidalgo}, S.~L., {Costa},
  E., \& {M{\'e}ndez}, R.~A. 2009, \apj, 705, 1260

\bibitem[{{No{\"e}l} {et~al.}(2007){No{\"e}l}, {Gallart}, {Costa}, \&
  {M{\'e}ndez}}]{NG07}
{No{\"e}l}, N.~E.~D., {Gallart}, C., {Costa}, E., \& {M{\'e}ndez}, R.~A. 2007,
  \aj, 133, 2037

\bibitem[{{Olsen} \& {Salyk}(2002)}]{os02}
{Olsen}, K.~A.~G. \& {Salyk}, C. 2002, \aj, 124, 2045

\bibitem [Pagel \& Tautvaisiene (1998)]{PT98}Pagel B. E. J., Tautvaisiene G., 1998, MNRAS, 299, 535\\

\bibitem[{{Paz} {et~al.}(2006){Paz}, {Lambas}, {Padilla}, \&
  {Merch{\'a}n}}]{p06}
{Paz}, D.~J., {Lambas}, D.~G., {Padilla}, N., \& {Merch{\'a}n}, M. 2006,
  \mnras, 366, 1503

\bibitem[{{Pejcha} \& {Stanek}(2009)}]{ps09}
{Pejcha}, O. \& {Stanek}, K.~Z. 2009, \apj, 704, 1730

\bibitem[{{Schlegel} {et~al.}(1998){Schlegel}, {Finkbeiner}, \& {Davis}}]{sc98}
{Schlegel}, D.~J., {Finkbeiner}, D.~P., \& {Davis}, M. 1998, \apj, 500, 525

\bibitem[Sarajedini (1999)]{Sar99} Sarajedini, A.1999, AJ, 118, 2321

\bibitem[{{Smith} {et~al.}(1992){Smith}, {Silbermann}, {Baird}, \&
  {Graham}}]{s92}
{Smith}, H.~A., {Silbermann}, N.~A., {Baird}, S.~R., \& {Graham}, J.~A. 1992,
  \aj, 104, 1430

\bibitem[{{Soszy{\~n}ski} {et~al.}(2010){Soszy{\~n}ski}, {Udalski},
  {Szyma{\~n}ski}, {Kubiak}, {Pietrzy{\~n}ski}, {Wyrzykowski}, {Ulaczyk}, \&
  {Poleski}}]{s10}
{Soszy{\~n}ski}, I., {Udalski}, A., {Szyma{\~n}ski}, M.~K., {Kubiak}, J.,
  {Pietrzy{\~n}ski}, G., {Wyrzykowski}, {\L}., {Ulaczyk}, K., \& {Poleski}, R.
  2010, Acta Astron, 60, 165

\bibitem[{{Soszy{\~n}ski} {et~al.}(2002){Soszy{\~n}ski}, {Udalski}, {Szymanski},
  {Kubiak}, {Pietrzynski}, {Wozniak}, {Zebrun}, {Szewczyk}, \&
  {Wyrzykowski}}]{s02}
{Soszy{\~n}ski}, I., {Udalski}, A., {Szymanski}, M., {Kubiak}, M., {Pietrzynski},
  G., {Wozniak}, P., {Zebrun}, K., {Szewczyk}, O., \& {Wyrzykowski}, L. 2002,
  Acta Astron, 52, 369

\bibitem[{{Stanek} {et~al.}(1998){Stanek}, {Zaritsky}, \& {Harris}}]{s98}
{Stanek}, K.~Z., {Zaritsky}, D., \& {Harris}, J. 1998, \apjl, 500, L141

\bibitem[{{Stanimirovi{\'c}} {et~al.}(2004){Stanimirovi{\'c}},
  {Staveley-Smith}, \& {Jones}}]{s04}
{Stanimirovi{\'c}}, S., {Staveley-Smith}, L., \& {Jones}, P.~A. 2004, \apj,
  604, 176

\bibitem[{{Subramaniam}(2003)}]{s03}
{Subramaniam}, A. 2003, \apjl, 598, L19

\bibitem[{{Subramaniam}(2005)}]{s05}
---. 2005, \aap, 430, 421

\bibitem[{{Subramaniam}(2006)}]{s06}
---. 2006, \aap, 449, 101

\bibitem[{{Subramanian} \& {Subramaniam}(2009)}]{ss09}
{Subramanian}, S. \& {Subramaniam}, A. 2009, \aap, 496, 399

\bibitem[{{Subramanian} \& {Subramaniam}(2010)}]{ss10}
---. 2010, \aap, 520, A24+

\bibitem[{{Suntzeff} {et~al.}(1986){Suntzeff}, {Friel}, {Klemola}, {Kraft}, \&
  {Graham}}]{s86}
{Suntzeff}, N.~B., {Friel}, E., {Klemola}, A., {Kraft}, R.~P., \& {Graham},
  J.~A. 1986, \aj, 91, 275

\bibitem[Tosi et al (2008)]{T08}Tosi, M., Gallagher, J., Sabbi, E., et al. 2008, 
IAUS, 255, 381

\bibitem[{{Tsujimoto} \& {Bekki}(2009)}]{t09}
{Tsujimoto}, T. \& {Bekki}, K. 2009, \apjl, 700, L69

\bibitem[{{Udalski}(1998)}]{u98}
{Udalski}, A. 1998, Acta Astron, 48, 383

\bibitem[{{Udalski} {et~al.}(2008){Udalski}, {Soszy{\'n}ski}, {Szyma{\'n}ski},
  {Kubiak}, {Pietrzy{\'n}ski}, {Wyrzykowski}, {Szewczyk}, {Ulaczyk}, \&
  {Poleski}}]{u08}
{Udalski}, A., {Soszy{\'n}ski}, I., {Szyma{\'n}ski}, M.~K., {Kubiak}, M.,
  {Pietrzy{\'n}ski}, G., {Wyrzykowski}, {\L}., {Szewczyk}, O., {Ulaczyk}, K.,
  \& {Poleski}, R. 2008, Acta Astron, 58, 329

\bibitem[{{van der Marel} \& {Cioni}(2001)}]{v01}
{van der Marel}, R.~P. \& {Cioni}, M. 2001, \aj, 122, 1807

\bibitem[{{van der Marel} {et~al.}(2009){van der Marel}, {Kallivayalil}, \&
 {Besla}}]{v09}
{van der Marel}, R.~P., {Kallivayalil}, N., \& {Besla}, G. 2009, 
The Magellanic System: Stars, Gas and Galaxies,, ed. {J.~T.~van Loon \& J.~M.~Oliveira},
 IAU Symposium, 256, 81



\bibitem[{{Weinberg} \& {Nikolaev}(2001)}]{wn01}
{Weinberg}, M.~D. \& {Nikolaev}, S. 2001, \apj, 548, 712


\bibitem[{{Westerlund}(1997)}]{w97}
Westerlund, B.E. 1997, The Magellanic Clouds, Cambridge (Cambridge Univ.Press)

\bibitem[{{Zaritsky} {et~al.}(2000){Zaritsky}, {Harris}, {Grebel}, \&
  {Thompson}}]{z00}
{Zaritsky}, D., {Harris}, J., {Grebel}, E.~K., \& {Thompson}, I.~B. 2000,
  \apjl, 534, L53

\end{thebibliography}
\end{document}